\documentclass[aps, 
prl,
twocolumn,
superscriptaddress,
showpacs,preprintnumbers,amsmath,amssymb,
longbibliography,floatfix,10pt]{revtex4-2}

\usepackage[
colorlinks=true, 
pdfstartview=FitV, 
linkcolor=blue, 
citecolor=magenta, 
urlcolor=blue, 
bookmarks=true,
bookmarksnumbered=true,
pdftitle={},
pdfauthor={}
]{hyperref}

\usepackage[dvips]{graphicx}
\usepackage{bm,latexsym,amsmath,amssymb,amsfonts,mathrsfs,textcomp}
\usepackage{color}
\input{colordvi.tex}
\usepackage{comment}
\usepackage{simpler-wick}
\usepackage{tikz}
\usepackage[compat=1.1.0]{tikz-feynhand}
\usepackage[normalem]{ulem}

\newcommand\sect[1]{{\it #1.}---}

\newcommand{\rmd}{\mathrm{d}} 

\newcommand{\pd}[2]{\frac{\partial #1}{\partial #2}}

\newcommand{\for}{\textrm{for}}

\newcommand{\ini}{\textrm{ini}}
\newcommand{\eq}{\textrm{eq}}

\newcommand{\att}{\textrm{att}}
\newcommand{\NS}{\textrm{NS}}
\newcommand{\eff}{\textrm{eff}}

\newcommand{\bx}{\bm{x}}
\newcommand{\by}{\bm{y}}

\newcommand{\bzero}{\bm{0}}

\newcommand{\calB}{\mathcal{B}}
\newcommand{\calC}{\mathcal{C}}

\newcommand{\calE}{\mathcal{E}}

\newcommand{\calH}{\mathcal{H}}

\newcommand{\calN}{\mathcal{N}}

\newcommand{\calP}{\mathcal{P}}

\newcommand{\calR}{\mathcal{R}}
\newcommand{\calS}{\mathcal{S}}

\newcommand{\braket}[1]{\langle {#1} \rangle}

\begin{document}

\title{Hydrodynamic attractor in ultracold atoms}

\author{Keisuke Fujii}
\email{fujii@phys.sci.isct.ac.jp}
\affiliation{Department of Physics, The University of Tokyo, Hongo, Bunkyo-ku, Tokyo 113-0033, Japan}
\affiliation{Institut f\"{u}r Theoretische Physik, Universit\"{a}t Heidelberg, Philosophenweg 19, 69120 Heidelberg, Germany}
\affiliation{Department of Physics, Institute of Science Tokyo, Ookayama, Meguro-ku, Tokyo 152-8551, Japan}

\author{Tilman Enss}
\affiliation{Institut f\"{u}r Theoretische Physik, Universit\"{a}t Heidelberg, Philosophenweg 19, 69120 Heidelberg, Germany}

\begin{abstract}
The hydrodynamic attractor is a concept that describes universal equilibration behavior in which systems lose microscopic details before hydrodynamics becomes applicable.
We propose a setup to observe hydrodynamic attractors in ultracold atomic gases, taking advantage of the fact that driving the two-body $s$-wave scattering length causes phenomena equivalent to isotropic fluid expansions.
We specifically consider two-component fermions with contact interactions in three dimensions and discuss their dynamics under a power-law drive of the scattering length in a uniform system.
By explicit computation, we derive a hydrodynamic relaxation model.
We analytically solve their dynamics and find the hydrodynamic attractor solution.
Our proposed method using the scattering length drive is applicable to a wide range of ultracold atomic systems, and our results establish these as a new platform for exploring hydrodynamic attractors.
\end{abstract}

\maketitle

\sect{Introduction}%
\label{sec:intro}%
Hydrodynamics universally describes the space-time evolution of charge densities of systems close to thermal equilibrium~\cite{Landau-Lifshitz:fluid}.
Hydrodynamic equations do not depend on the microscopic details of systems, and are applicable to broad areas of physics from condensed matter~\cite{Chaikin:2000} to high energy physics~\cite{Rischke:1996,Vogelsberger:2014}.
Its universality is based on the coarse graining of microscopic elements into macroscopic fluid cells.
For time scales sufficiently longer than relaxation times, one can describe the dynamics of the fluid cells in terms of charge densities and their derivatives, and systematically write down hydrodynamic equations based on gradient expansions.
The hydrodynamic equations at leading order in the expansion describe the ideal fluid, those up to the first order describe the Navier-Stokes fluid, and further higher-order corrections can be found as needed.

However, recent high-energy heavy-ion collision experiments reported that their initial dynamics immediately after the impact of two relativistic nuclei can be described hydrodynamically, even though it is far from equilibrium~\cite{CMS:2010ifv,ALICE:2012eyl,ATLAS:2012cix,PHENIX:2018lia,STAR:2022pfn}.
This ``unreasonable'' effectiveness of hydrodynamics triggered a reconsideration of its applicability, and suggests the existence of 
non-equilibrium universal attractors to hydrodynamics, which cannot be captured within naive gradient expansions in hydrodynamics~\cite{Heller:2015}.
Such attractors, called hydrodynamic attractors, have been actively studied and found from various microscopic theories such as hydrodynamics, kinetic theory~\cite{Kurkela:2019}, and holography~\cite{Heller:2015,Romatschke:2018, kurkela:2020} (see review papers~\cite{Romatschke:2017,Florkowski:2018,Romatschke:2019,Berges:2021,Soloviev:2022}).

The hydrodynamic attractor is also relevant for small systems whose typical time and length scales are comparable to their relaxation times and mean free paths.
Recently, in cold atoms, such small systems were realized with the development of experimental techniques.
Relaxation times of strongly interacting Fermi gases were measured through initial state preparation and time-resolved measurements, both in trapped gases~\cite{Vogt:2012,Brewer:2015} and in uniform systems~\cite{Patel:2020,Li:2022,Li:2024}.
In particular, the detailed collective dynamics of a few strongly correlated fermions were measured~\cite{Brandstetter:2023}.
These experiments make cold atomic systems an important new research platform for hydrodynamic attractors.

\begin{figure}[t]
 \centering
\includegraphics[width=\linewidth]{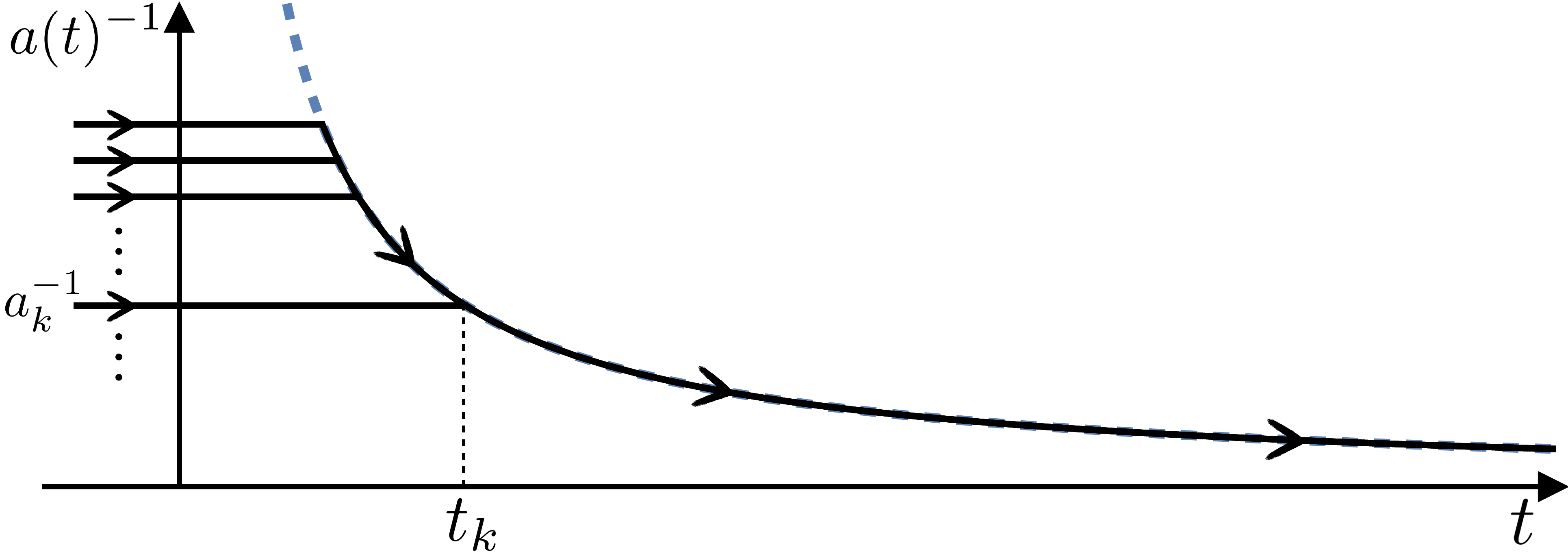}
\caption{Protocol for driving the scattering length to realize the hydrodynamic attractor in a uniform system without fluid velocities.
The scattering length is kept at a constant value $a_k$ up to time $t=t_k$ and then approaches the unitary limit $a^{-1}(t)\to0$ asymptotically with a power law of time; see Eq.~\eqref{eq:protocol}. 
To probe various initial conditions, the initial scattering length $a_k$ and time $t_k$ are varied while keeping $\tilde{a}=a_{k}(\tau_{\zeta}/t_{k})^{\alpha}$ fixed for $k=1,2,3,\ldots$.
Once the drive has started for $t>t_k$, the scattering length follows a single curve.
}
 \label{fig:protocol}
\end{figure}

In this Letter, we propose a setup to study hydrodynamic attractors in cold atoms.
We consider a two-component Fermi gas at any temperature in three dimensions whose short-range interaction is fully characterized by the two-body $s$-wave scattering length and discuss its hydrodynamic behavior when the scattering length is changed in time.
In this system, the time variation of the scattering length at fixed volume leads to phenomena equivalent to isotropic fluid expansions of the gas because there are no other intrinsic reference scales~\cite{Fujii:2018}.
In other words, by temporally varying the scattering length in a uniform state without fluid velocities, one can arbitrarily drive the system out of equilibrium, equivalent to isotropic fluid expansion, while the fluid remains uniform and at rest.
Taking advantage of this equivalence, we show that hydrodynamic attractors can be explored in cold atomic systems by driving the scattering length to the strongly interacting, unitary limit over time, as schematically depicted in Fig.~\ref{fig:protocol}.
We note that our proposed method is applicable not only to fermions near unitarity, but to a wide range of ultracold atomic systems whose interaction is characterized by the $s$-wave scattering length alone, such as fermions in the entire BCS-BEC crossover and repulsively interacting Bose gases.

\sect{Bulk viscosity}%
\label{sec:bulk-viscosity}%
Let us start with a brief review of the bulk viscosity, which characterizes dissipation in isotropic fluid expansion.
According to linear-response theory, the complex bulk viscosity $\zeta(\omega)$ in the normal phase [replaced by $\zeta_2(\omega)$ in the superfluid phase] at frequency $\omega$ is provided by~\cite{Mori:1962,Luttinger:1964,Bradlyn:2012,Fujii:2020}
\begin{align}
\zeta(\omega)=\frac{\calR_{\Pi\Pi}(\omega+i0^{+})-\calR_{\Pi\Pi}(i0^{+})}{i(\omega+i0^{+})}, \label{eq:complex-bulk-viscosity}
\end{align}
where $\calR_{\Pi\Pi}(w)\equiv i\int^{\infty}_{0}\rmd t\int \rmd \bx\,e^{iwt}\langle[\hat{\Pi}(t,\bx),\hat{\Pi}(0,\bzero)]\rangle$ is the response function of pressure fluctuations. These are given by the modified trace of the stress tensor, $\hat{\Pi}\equiv \hat{p}-\left(\partial p/\partial \calN\right)_{\calE}\hat{\calN}-\left(\partial p/\partial \calE\right)_{\calN}\hat{\calH}$, and $\langle\cdots\rangle$ represents the grand canonical expectation value.
The pressure operator $\hat{p}=\sum_{i}\hat{\Pi}_{ii}/3$ is defined by the trace of the stress tensor operator $\hat{\Pi}_{ij}$, while $\hat{\calN}$ denotes the number density operator and $\hat{\calH}$ the Hamiltonian density operator, with $p=\braket{\hat{p}}$, $\calN=\braket{\hat{\calN}}$, and $\calE=\braket{\hat{\calH}}$.

Two-component fermions with attractive zero-range interaction are described by the Hamiltonian density~\cite{Zwerger:2012}
\begin{align}
\hat{\calH}
=\sum_{\sigma=\uparrow,\downarrow}\hat{\psi}^{\dagger}_{\sigma}\frac{-\nabla^{2}}{2m}\hat{\psi}_{\sigma}
+g_{0}\hat{\psi}^{\dagger}_{\uparrow}\hat{\psi}^{\dagger}_{\downarrow}\hat{\psi}_{\downarrow}\hat{\psi}_{\uparrow},
\end{align}
where $\hat{\psi}^{\dagger}_{\sigma}$ is the creation operator for spin-$\sigma$ fermions (we set $\hbar=1$).
In dimensional regularization, the coupling constant $g_{0}$ is related to the scattering length $a$ via $mg_0=4\pi a$.
In this system, the pressure operator satisfies~\cite{Suppl}
\begin{align}
\hat{p}=\frac{2}{3}\hat{\calH}+\frac{\hat{\calC}}{12\pi m a},\label{eq:pressure-rel}
\end{align}
up to irrelevant total derivatives, where the contact density operator $\hat{\calC}\equiv (mg_{0})^{2}\hat{\psi}^{\dagger}_{\uparrow}\hat{\psi}^{\dagger}_{\downarrow}\hat{\psi}_{\downarrow}\hat{\psi}_{\uparrow}$ describes short-range pair correlations~\cite{Braaten:2008}.
This operator identity is the nonrelativistic counterpart of the condition of tracelessness due to conformal invariance, and the second term on the right-hand side measures the breaking of the conformal symmetry.
If the interaction strength is tuned to the unitary limit, i.e., $|a|=\infty$, the second term in Eq.~\eqref{eq:pressure-rel} vanishes, and accordingly the equation of state obeys $p=2\calE/3$.
In the unitary limit, therefore, the modified trace $\hat{\Pi}$ becomes zero, and the complex bulk viscosity $\zeta(\omega)$ vanishes identically~\cite{Nishida:2007,Son:2007}.

The operator $\hat{\Pi}$ can be expressed by the contact density operator up to conserved density operators using Eq.~\eqref{eq:pressure-rel}, so that the complex bulk viscosity is expressed in terms of the response function of the contact density:
\begin{align}
 \zeta(\omega)
 =\frac{1}{(12\pi m a)^2}
 \frac{\calR_{\calC\calC}(\omega+i0^{+})-\calR_{\calC\calC}(i0^{+})}{i(\omega+i0^{+})},
\end{align}
where the commutator of the Hamiltonian and particle number operators with any operator in the grand canonical average can safely be dropped.
The response function of the contact density captures how the contact density at time $t$ changes in response to a variation of the scattering length at time $t=0$ as
\begin{align}
\label{eq:contactcorrel}
i\langle[\hat{\calC}(t,\bx),\hat{\calC}(0,\bzero)]\rangle
=-4\pi m\biggl(\pd{\langle\calC(t,\bx)\rangle}{a(0,\bzero)^{-1}}\biggr)_{S,N}
\end{align}
at constant entropy and particle number~\cite{Enss:2019, Fujii:2020}.
Thus, both isotropic expansion and changes in scattering length can be used to measure $\zeta(\omega)$.

The full frequency dependence of the bulk viscosity has been computed numerically for the strongly correlated fluid around unitarity in the quantum degenerate regime above the superfluid transition \cite{Enss:2019} and analytically at high temperature \cite{Dusling:2013,Nishida:2019,Enss:2019,Hofmann:2020,Fujii:2023}. The Luttinger-Ward approach is based on dressed fermions which decay by scattering off molecules, and dressed molecules which decay by dissociation into pairs of separate fermions. This decay of molecules provides a large contribution to the bulk viscosity.
At low frequencies, the numerical $\zeta(\omega)$ is accurately represented by the Drude form
\begin{align}
\label{eq:drude}
\zeta(\omega)=\frac{i\chi}{\omega+i\tau_{\zeta}^{-1}},
\end{align}
where $\tau_{\zeta}$ is the relaxation time for the bulk viscosity that captures the details of many-body dynamics. The sum rule $\chi=(1/\pi)\int_{-\infty}^\infty d\omega\,\zeta(\omega)=(1+2/d)p-\calN(\partial p/\partial \calN)_{\mathcal S}$ is a thermodynamic property \cite{Landau-Lifshitz:fluid,Taylor:2012}.
Both $\tau_\zeta$ and $\chi$ incorporate all microscopic details, such as correlations between bound or virtual pairs, and the scattering rate $\tau_\zeta^{-1}\propto k_BT$ is found to be comparable to temperature.
At frequencies $\omega\gtrsim \varepsilon_F$ larger than the Fermi energy the complex bulk viscosity has a high-frequency tail proportional to $\omega^{-3/2}$ due to the short-distance singularity of the contact interaction~\cite{Hofmann:2011,Goldberger:2012}, but this is irrelevant for the intermediate- and long-time dynamics that we focus on.
While we have arrived at the Drude form \eqref{eq:drude} by explicit numerical computation, it is valid more generally in the BCS-BEC crossover and can be derived systematically in the memory function formalism~\cite{Gotze:1972}; this has recently been successfully applied to systems without well-defined quasiparticles~\cite{Hartnoll:2018}, such as the unitary Fermi gas near the critical temperature~\cite{Frank:2020}.
Furthermore, the Drude form is fundamental to providing a systematic and unified description for deriving the hydrodynamic equations of motion~\cite{Forster}.

\sect{Time-dependent scattering length in hydrodynamics}%
\label{sec:hydrodynamics}%
Let us consider hydrodynamically a specific situation in which the scattering length $a(t)$ is varied over time in a uniform system without fluid velocities.
Then, the energy density is produced at the rate of
\begin{align}
\dot{\calE}(t)=\frac{\calC(t)}{4\pi m a(t)^{2}}\dot{a}(t), \label{eq:dynamic-sweep}
\end{align}
which is known as the dynamic sweep theorem~\cite{Tan:2008a,Tan:2008b,Tan:2008c}.
Here, the contact density expectation value $\calC(t)$ is divided into two parts as
\begin{align}
\calC(t)=\calC_{\eq}[a(t)]+12\pi m a(t)\pi(t),
\end{align}
where the first term gives the instantaneous contact density determined in thermal equilibrium with the scattering length $a(t)$, and $\pi(t)$ in the second term describes the dissipative correction to the first term in hydrodynamics.
In local thermal equilibrium, the thermodynamic densities satisfy the instantaneous version of the pressure relation $p(t)=2\calE(t)/3+\calC_{\eq}[a(t)]/12\pi m a(t)$.
By comparing this instantaneous pressure relation with Eq.~\eqref{eq:pressure-rel}, one can view $\pi(t)$ hydrodynamically as the deviation of the pressure operator from its equilibrium expectation.

The Drude form \eqref{eq:drude} corresponds to exponential relaxation $\zeta(t)=\chi \exp(-t/\tau_\zeta) \Theta(t)$. For small-amplitude changes of $a(t)$, even when they are rapid, linear-response theory \cite{Forster} for Eq.~\eqref{eq:contactcorrel} yields the memory integral $\pi(t)=-\int_{-\infty}^t dt'\, \zeta(t-t') V_a(t')$ for drive $V_a(t)$ below. By time derivative, we obtain the equation of motion
\begin{align}
\tau_{\zeta}\dot{\pi}(t)+\pi(t)=-\zeta[a(t)]V_a(t),
\label{eq:MIS-type}
\end{align}
with relaxation time $\tau_{\zeta}=\zeta(\omega\to0)/\chi$.
Here, $\zeta[a]$ is the static bulk viscosity coefficient $\lim_{\omega\to 0}\zeta(\omega)$ for given scattering length $a$, while
$V_a(t)$ is the bulk strain rate tensor modified by the time-dependent scattering length and is given by~\footnote{In Ref.~\cite{Fujii:2018}, the bulk strain rate tensor with a space-time dependent scattering length is generally derived as $V_{a}(t,\bx)=\nabla\cdot\bm{v}(t,\bx)-3[\partial_t a(t,\bx)+\bm{v}(t,\bx)\cdot\nabla a(t,\bx)]/a(t,\bx)$, where $\bm{v}(t,\bx)$ is the fluid velocity and $a(t,\bx)$ is the space-time dependent scattering length.
In uniform systems without fluid velocities, the bulk strain rate tensor is given only by the time derivative term of the scattering length.}
\begin{align}
V_{a}(t)\equiv
-3\dot a(t)/a(t).
\end{align}
Whereas the bulk strain rate tensor is given by the divergence of fluid velocities in ordinary hydrodynamics, $V_a(t)$ also has the time derivative term of the scattering length.
This is a consequence of the equivalence between isotropic fluid expansion and temporal contraction of the scattering length. Remarkably, one can thus drive the fluid out of equilibrium by the scattering length and observe how it relaxes back toward equilibrium via the contact, without moving parts in a uniform system.

If we take the long-time limit $t/\tau_{\zeta}\to\infty$, i.e., the limit of $\tau_{\zeta}\to 0$, in Eq.~\eqref{eq:MIS-type}, $\pi(t)$ is reduced to $\pi(t)|_{\NS}=-\zeta[a(t)]V_a(t)$, which gives the Navier-Stokes hydrodynamic result.
Furthermore, higher-order hydrodynamic corrections beyond the Navier-Stokes level can be obtained systematically by expanding $\pi(t)$ with respect to $\tau_{\zeta}$.
In relativistic systems, hydrodynamic equations with relaxation times in the form of Eq.~\eqref{eq:MIS-type} are commonly employed to preserve causality and are referred to as the M\"{u}ller-Israel-Stewart theory~\cite{Muller:1967,Israel:1976,Israel:1979}.

\sect{Hydrodynamic attractor}%
\label{sec:attractor}%
In order to realize the universal hydrodynamic attractor in the relaxation dynamics, we drive the scattering length so that the system is initially brought out of equilibrium and then gradually approaches thermal equilibrium.
By solving Eq.~\eqref{eq:MIS-type} with any such drive for a given functional form of $\zeta[a]$, one can find the hydrodynamic attractors.

To present analytical results, we now consider a specific class of drives that realize a hydrodynamic attractor.
We investigate the situation where the system asymptotically approaches the unitary limit by a power-law variation of the scattering length with exponent $\alpha>0$ as
\begin{align}
a_{k}(t)^{-1}=
\left\{\!
\begin{array}{ll}
a^{-1}_{k} & t<t_k, \\[0.8ex]
a^{-1}_{k}(t/t_k)^{-\alpha} & t>t_k,
\end{array}
\right.\ \  \for\ k=1,2,\ldots,
\label{eq:protocol}
\end{align}
where $a_{k}$ is a constant scattering length up to time $t_k$.
Here, $k$ is a subscript to distinguish the drives for different initial conditions.
To make the drive the same at long times, we fix $a_{k}(\tau_{\zeta}/t_{k})^{\alpha}=:\tilde{a}$, which gives the scattering length at the relaxation time in the power-law drive.
Therefore, at long times, all the driving protocols lie on a single curve that asymptotically approaches the unitary limit at a power $\alpha$, as shown in Fig.~\ref{fig:protocol}.

We suppose that the system is sufficiently close to the unitary limit to maintain the system in the linear response regime.
Near the unitary limit, the static bulk viscosity $\zeta[a]$ is proportional to $a^{-2}$~\cite{Dusling:2013}, so that we take an approximated form as $\zeta[a(t)]\simeq \zeta^{(2)}a(t)^{-2}$ with $\zeta^{(2)}$ being a constant.
Then, Eq.~\eqref{eq:MIS-type} for $t>t_k$ turns into
\begin{align}
\tau_{\zeta}\dot{\pi}(t)+\pi(t)
=3\zeta[\tilde{a}] \frac{\alpha\tau_{\zeta}^{2\alpha}}{t^{2\alpha+1}},
\label{eq:MIS-eq-power-law-drive}
\end{align}
and its analytical solution is found to be
\begin{align}
\pi(t)=\pi_{\att}(t)+\pi_{\ini}e^{-t/\tau_{\zeta}},
\label{eq:exact-sol}
\end{align}
where $\pi_{\ini}=-e^{t_k/\tau_{\zeta}}\pi_{\att}(t_k)$ is determined from the initial condition $\pi(t_k)=0$ because the system is in equilibrium with a constant scattering length up to time $t=t_{k}$.
Here, $\pi_{\att}(t)$ is given by
\begin{align}
\pi_{\att}(t)=\frac{3\zeta[\tilde{a}]\alpha}{\tau_{\zeta}}(-1)^{2\alpha+1}e^{-t/\tau_{\zeta}}\Gamma(-2\alpha,-t/\tau_{\zeta}),
\label{eq:attractor-sol}
\end{align}
with $\Gamma(s,x)$ being the incomplete Gamma function~%
\footnote{
The incomplete Gamma function $\Gamma(-s,-x)$ for $s>0$ and $x>0$ appears from 
 the integral $\int^{x}_{x_0}dt\, e^{t}t^{-s-1}=(-1)^{s+1}[\Gamma(-s,-x)-\Gamma(-s,-x_{0})]$ and formally describes the general solution independent of the initial conditions for the hydrodynamic equation~\eqref{eq:MIS-eq-power-law-drive}.
}.
The first term of Eq.~\eqref{eq:exact-sol}, $\pi_{\att}(t)$, depends on $a_k$ and $t_k$ not separately but only through the fixed parameter $\tilde{a}$, while the second term explicitly depends on $t_k$.
Therefore, $\pi_{\att}(t)$ is independent of the initial condition.

\begin{figure}[t]
 \centering
 \includegraphics[width=0.84\linewidth]{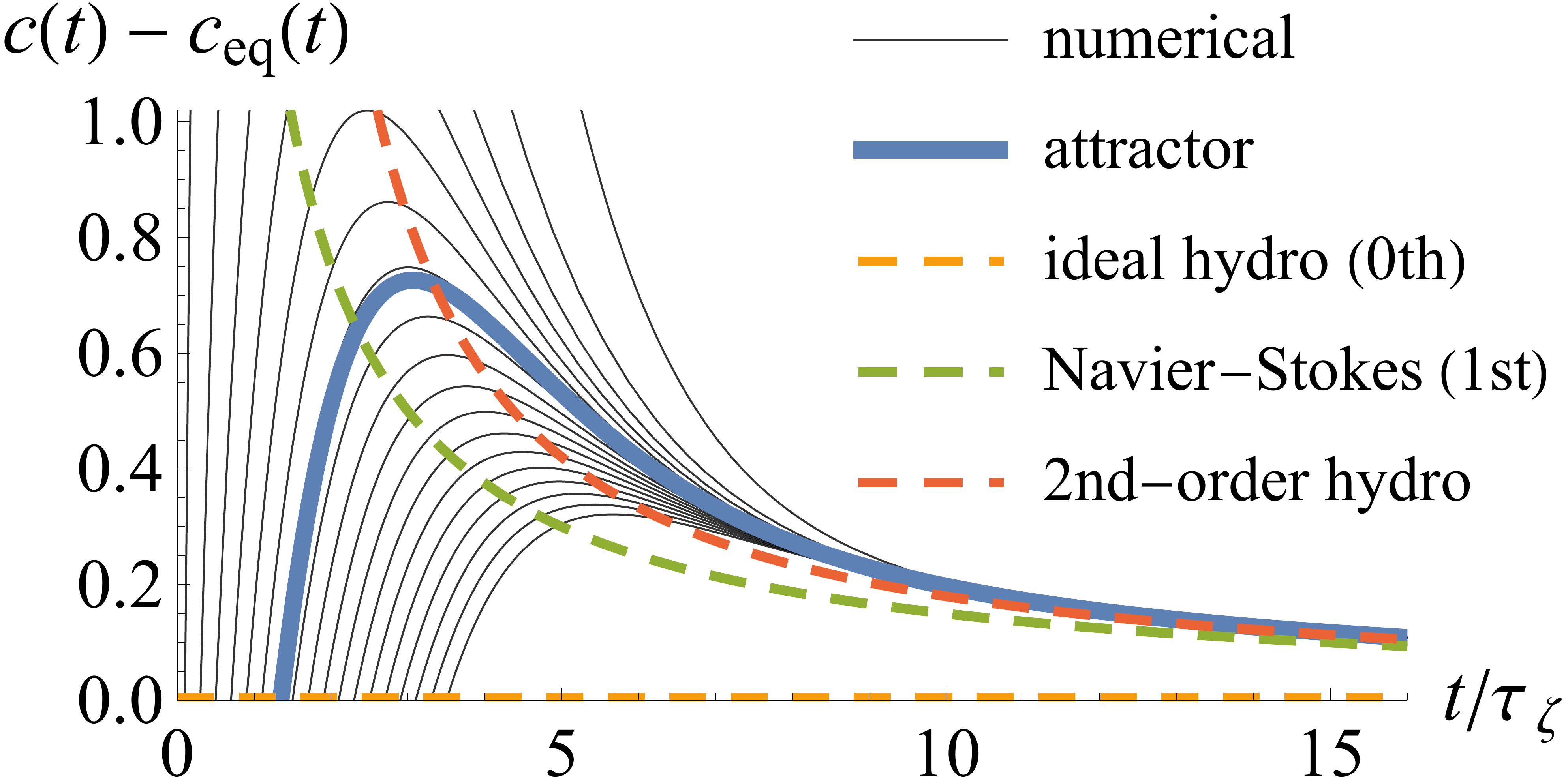}
\caption{
The hydrodynamic attractor solution (blue thick line) for the deviation of the dimensionless contact density from its equilibrium value, $c(t)-c_{\eq}[a(t)]\equiv(\calC(t)-\calC_{\eq}[a(t)])/(12\pi m a(t))\times (\tau_{\zeta}/\zeta[a(t)])$, as a function of $t/\tau_{\zeta}$ under power-law drive with $\alpha=1/2$. 
We also plot numerical solutions (black thin lines) of Eq.~\eqref{eq:exact-sol} for $t_k/\tau_{\zeta}=0.1,0.3,0.5,\ldots,3.5$ and hydrodynamic results of zeroth-order (orange dashed line), first-order (green dashed line), and second-order (red dashed line) from the expansion~\eqref{eq:expansion}.
The universal attractor solution corresponds to the numerical solution with $t_k/\tau_{\zeta}=1.347$.
}
 \label{fig:contact}
\end{figure}

Let us investigate the behavior of the solution $\pi(t)$ in the long-time limit, where the Navier-Stokes hydrodynamics have to be reproduced.
The first term $\pi_{\att}(t)$ is expanded for long times $t\gg \tau_{\zeta}$ as
\begin{align}
\pi_{\att}(t)
&=3\zeta[\tilde{a}]
\frac{\alpha\tau_{\zeta}^{2\alpha}}{t^{2\alpha+1}}
\biggl[
	1+(2\alpha+1)\frac{\tau_{\zeta}}{t}+O\left((\tau_{\zeta}/t)^{2}\right)
\biggr],
\label{eq:expansion}
\end{align}
where the exponential factor $e^{-t/\tau_{\zeta}}$ in Eq.~\eqref{eq:attractor-sol} is harmless in the expansion with respect to $t/\tau_{\zeta}$ due to the asymptotic behavior of the incomplete Gamma function, $\Gamma(-s,-z)= (-1/z)^{s+1}e^{z}/\Gamma(s+1)\sum_{n=0}^{\infty}\Gamma(s+n+1)z^{-n}$ for $|z|\to\infty$.
In this expansion, the first term coincides with the Navier-Stokes result, and the second term gives a second-order hydrodynamic correction.
In contrast, the initial condition in the second term in Eq.~\eqref{eq:exact-sol} has an exponential damping factor $e^{-(t-t_k)/\tau_{\zeta}}$, which cannot be expanded with respect to $\tau_{\zeta}/t$, and is called the non-hydrodynamic mode.

The key point is that the expansion~\eqref{eq:expansion} is asymptotic with convergence radius zero.
The coefficient of the $n\,$th-order term in the expansion is, indeed, proportional to $\Gamma(2\alpha+n+1)$ and diverges factorially.
This factorial divergence makes the gradient expansion underlying hydrodynamics a divergent series and significantly less accurate for small $t/\tau_{\zeta}$.
On the other hand, $\pi_{\att}(t)$ itself, given analytically in Eq.~\eqref{eq:attractor-sol}, universally describes the system accurately, independent of the initial conditions, for $t/\tau_{\zeta} \gtrsim 1$ after the non-hydrodynamic mode has decayed exponentially.
Because the universality emerges before the time scale $t\gg \tau_\zeta$ at which hydrodynamics becomes accurate, $\pi_{\att}(t)$ is called \textit{the hydrodynamic attractor}.

Note that the expression~\eqref{eq:attractor-sol} has non-analytic contributions with respect to $\tau_\zeta/t$, while the expansion~\eqref{eq:expansion} does not retain such contributions.
For example, the attractor solution for $\alpha=1/2$ is expanded for short times $t\ll\tau_{\zeta}$ as 
\begin{align}
 \pi_{\att}(t)|_{\alpha=1/2}
 =\frac{3\zeta[\tilde{a}]}{2\tau_{\zeta}}
\biggl[
    -\frac{\tau_{\zeta}}{t}+\gamma+\ln\frac{t}{\tau_\zeta}
   +O(t/\tau_{\zeta})
\biggr],~\label{eq:short-time-expansion}
\end{align}
with $\gamma=0.5772\ldots$ being Euler's constant.
Using the Borel summation, we can find the exact attractor solution~\eqref{eq:attractor-sol} from the expansion~\eqref{eq:expansion}~\cite{Suppl}.
The Borel summation for effective theories is one of the common approaches to finding the hydrodynamic attractors~\cite{Heller:2015}.

\sect{Contact density}%
\label{sec:densities}%
Figure~\ref{fig:contact} plots the deviation of the dimensionless contact density $c(t)\equiv \calC(t)/(12\pi m a(t))\times (\tau_{\zeta}/\zeta[a(t)])$ from its instantaneous value under the drive of the scattering length with a power $\alpha=1/2$.
Here, the deviation of the dimensionless contact density is equal to $\pi(t)\tau_{\zeta}/\zeta[a(t)]$, so that it corresponds to the pressure deviation from its equilibrium value and does not depend on the fixed parameter $\tilde{a}$.
Numerical solutions of Eq.~\eqref{eq:MIS-eq-power-law-drive} for $t_k/{\tau_{\zeta}}=0.1,0.3,\ldots,3.5$ (black thin lines) remain zero (their equilibrium values) until the start of driving at $t=t_k$, and then take positive values as shown in Fig.~\ref{fig:contact}.
Since the bulk strain rate tensor becomes smaller as time increases, the system approaches thermal equilibrium, and thus the deviation tends toward zero.
Importantly, the numerical solutions converge rapidly to the universal attractor solution (blue thick line) before being reduced to the hydrodynamic solutions (dashed lines).

The attractor behavior can be probed in current cold-atom experiments, e.g., in a two-component Fermi gas of $^{40}$K atoms with Fermi energy $E_F\sim h\times 20$\,kHz at $T/T_F=0.25$.
The effect is expected to be large near unitarity, where the bulk viscous relaxation time is predicted as $\tau_{\zeta}\sim 0.7\hbar/k_B T\sim 0.15$\,ms~\cite{Enss:2019}.
Let us consider the power-law protocol with $\alpha=1/2$ and $1/k_F\tilde{a}=0.5$.
As starting times of the ramp, one can choose $t_k/\tau_{\zeta}=0.5,1,2$;
the maximum rate of change of $1/k_F a$ is then $\sim 4$\,kHz, well within experimental capabilities near the narrow Feshbach resonance at 202\,G. 
The attractor is expected to occur in the time window $t\sim $(5--10)$\times\tau_{\zeta}\sim$ 0.75--1.5\,ms.
It will be visible in the time-dependent contact, which has already been measured with 0.1\,ms time resolution~\cite{Bardon:2014,Luciuk:2017}.

From the perspective of the Navier-Stokes solution, the attractor appears to result from an effective bulk viscosity coefficient that varies at short times before it approaches its equilibrium value at long times.  The effective bulk viscosity coefficient $\zeta^{(\alpha)}_{\eff}[a(t)]$ can be defined for $t>t_k$ by representing the hydrodynamic attractor solution in the form of the Navier-Stokes result~\cite{Romatschke:2018}:
\begin{align}
 \pi_{\att}(t)=3\zeta^{(\alpha)}_{\eff}[a(t)]\frac{\alpha}{t}.
\end{align}
From the explicit form of the attractor solution~\eqref{eq:attractor-sol}, we find
\begin{align}
\frac{\zeta^{(\alpha)}_{\eff}[a(t)]}{\zeta[a(t)]}
= \biggl(-\frac{t}{\tau_{\zeta}}\biggr)^{2\alpha+1}e^{-t/\tau_{\zeta}}\Gamma(-2\alpha,-t/\tau_{\zeta}).
\label{eq:ratio-effViscosity}
\end{align}
The effective viscosity $\zeta^{(\alpha)}_{\eff}[a(t)]$ has non-analytic contributions in the expansion with respect to $t/\tau_{\zeta}$, analogous to the short-time expansion of $\pi_{\att}(t)$ in Eq.~\eqref{eq:short-time-expansion}.

\begin{figure}[t]
 \centering
 \includegraphics[width=0.94\linewidth]{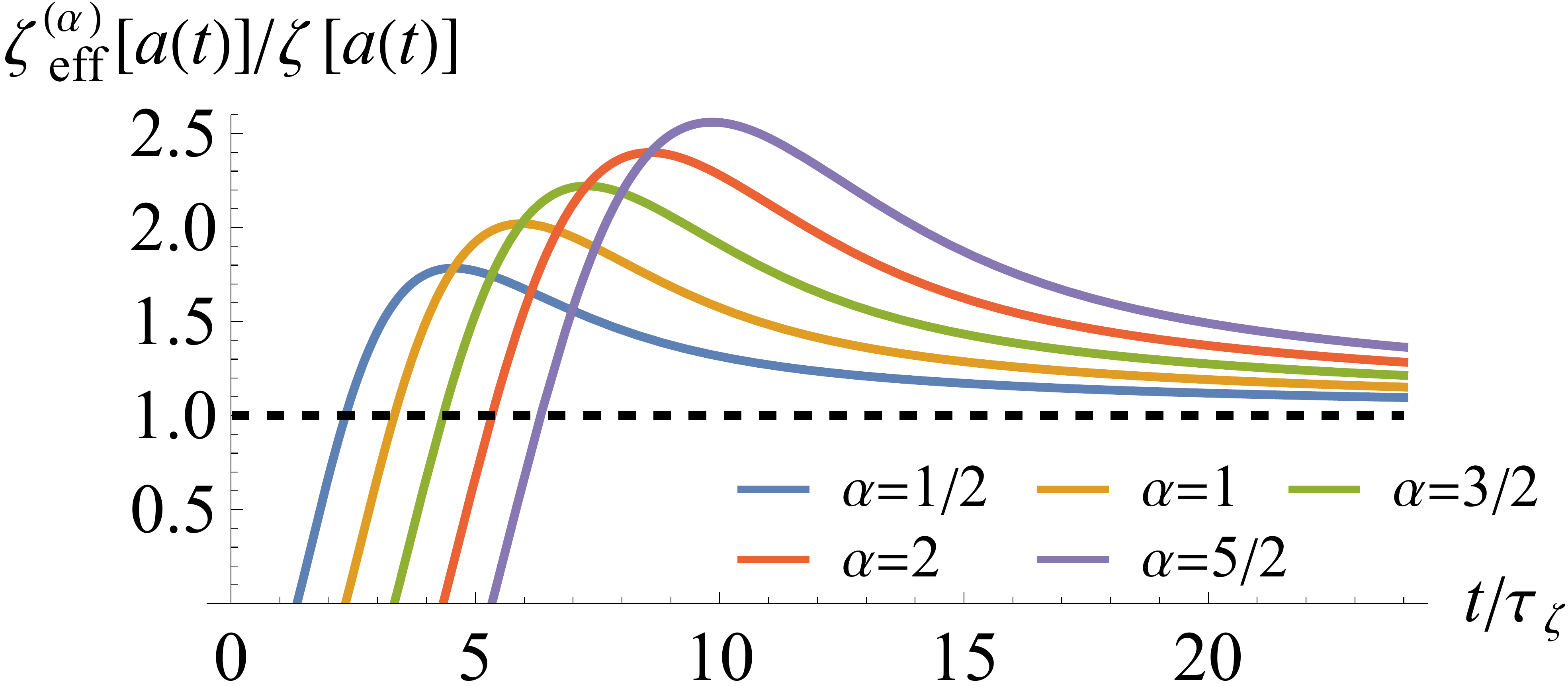}
\caption{
The effective, time-dependent bulk viscosity coefficients reproduce the attractor solution when inserted into the Navier-Stokes dynamics~\eqref{eq:ratio-effViscosity}; shown here for different drive exponents $\alpha$.
}
 \label{fig:effViscosity}
\end{figure}

Figure~\ref{fig:effViscosity} plots the effective bulk viscosity as a function of $t/\tau_{\zeta}$ for various powers.
Since the ratio~\eqref{eq:ratio-effViscosity} measures the deviation of the attractor solution from the Navier-Stokes result, it asymptotically approaches unity in the long-time limit.
The ratio has a peak for intermediate times and is suppressed for short times.
These tendencies are generally found in the effective viscosity coefficients defined by the hydrodynamic attractor~\cite{Romatschke:2018}.

\sect{Discussion and outlook}%
\label{sec:Discussion}%
In this Letter, we have proposed a driving protocol of the scattering length to observe the hydrodynamic attractor in cold atoms; this protocol is given by Eq.~\eqref{eq:protocol} and schematically depicted in Fig.~\ref{fig:protocol}. 
Employing the hydrodynamic equation with the relaxation time [Eq.~\eqref{eq:MIS-type}], we have analytically found the attractor solution and the non-hydrodynamic mode as Eq.~\eqref{eq:exact-sol}.
The dynamics of the contact density in our protocol is plotted in Fig.~\ref{fig:contact}, where the solutions from different initial conditions converge to the attractor solution already before the time scale where the hydrodynamics become relevant.
Similarly, one can find the dynamics of other thermodynamic variables such as the energy density by integrating Eq.~\eqref{eq:dynamic-sweep}.
The deviation between the attractor solution and the Navier-Stokes solution is measured as the ratio of the effective viscosity coefficient in Eq.~\eqref{eq:ratio-effViscosity} and Fig.~\ref{fig:effViscosity}.
Since we can choose the power $\alpha$ arbitrarily, these results would be useful for measuring the bulk viscosity coefficient $\zeta[a]$ near the unitary limit, or more precisely $\zeta^{(2)}$.

Since our analysis neglects the high-frequency tail of the correlation function due to the singularity of the short-range interaction, the short-time behavior of our results is not exact.
However, the details near the initial time immediately disappear as non-hydrodynamic modes, and the dynamics universally follow the attractor solution for intermediate and long times.
On the other hand, the power-law drive of the scattering length at a power $\alpha=1/2$ leads to another universality in the short-time dynamics~\cite{Qi:2021} due to the scale invariance ($t\to \lambda^2 t$ and $\bx\to\lambda \bx$) of the zero-density nonrelativistic system.
Moreover, the nonrelativistic scale invariance characterizes the stability of the system near the unitary limit under the power-law drive~\cite{Maki:2022}.
It will be worthwhile to explore how this short-time universality caused by the scale invariance and the intermediate- and long-time universality of the hydrodynamic attractor can be connected.

It is also worthwhile to remark that our proposed protocol is analogous to a relativistic hydrodynamic attractor in a Hubble expansion~\cite{Du:2021}.
This is because the time variation of the scattering length corresponds to the time variation of the determinant of the spatial metric~\footnote{
The bulk strain rate tensor coupled to a spatial metric $g_{ij}(t,\bx)$ is generally given as $V_{a}(t,\bx)=\sum_{k=1}^{3}\nabla_{k}v^{k}(t,\bx)+\partial_{t}\ln\sqrt{g(t,\bx)}-3[\partial_{t} a(t,\bx)+v^{k}(t,\bx)\partial_{k} a(t,\bx)]/a(t,\bx)$, where $v^{k}(t,\bx)$, $g(t,\bx)$, and $a(t,\bx)$ are fluid velocities, the determinant of the spatial metric, and the spacetime-dependent scattering length, respectively~\cite{Son:2007,Fujii:2018}.
Here, $\nabla_{k}$ is the covariant derivative with respect to the spatial metric $g_{ij}(t,\bx)$}.
Therefore, it is significant to explore hydrodynamic attractors in cold atom systems, where the initial-time dynamics can be directly observed, unlike in heavy-ion collision experiments. Specifically, the scattering length can be tuned by changes in the magnetic field~\cite{Chin:2010}, while the contact dynamics have been measured with high resolution in time~\cite{Bardon:2014, Luciuk:2017}.
In particular, our proposed protocol allows the choice of arbitrary power of the driving and, moreover, allows various initial states to be realized in a well-controlled manner.

\smallskip
\begin{acknowledgments}
The authors acknowledge useful discussions with J.~Maki, A.~Mazeliauskas, J.~H.~Thywissen, W.~van der Schee, and U.~Wiedemann,
as well as participants at the EMMI Rapid Reaction Task Force ``Deciphering many-body dynamics in mesoscopic quantum gases.'' 
This work is supported by the Deutsche Forschungsgemeinschaft (DFG, German Research Foundation), project-ID 273811115 (SFB1225 ISOQUANT), and under Germany's Excellence Strategy EXC2181/1-390900948 (the Heidelberg STRUCTURES Excellence Cluster).
K.F. is supported by JSPS KAKENHI Grant Number JP24KJ0062.
\end{acknowledgments}

\bibliography{hydro-attractor}

\begin{thebibliography}{62}%
\makeatletter
\providecommand \@ifxundefined [1]{%
 \@ifx{#1\undefined}
}%
\providecommand \@ifnum [1]{%
 \ifnum #1\expandafter \@firstoftwo
 \else \expandafter \@secondoftwo
 \fi
}%
\providecommand \@ifx [1]{%
 \ifx #1\expandafter \@firstoftwo
 \else \expandafter \@secondoftwo
 \fi
}%
\providecommand \natexlab [1]{#1}%
\providecommand \enquote  [1]{``#1''}%
\providecommand \bibnamefont  [1]{#1}%
\providecommand \bibfnamefont [1]{#1}%
\providecommand \citenamefont [1]{#1}%
\providecommand \href@noop [0]{\@secondoftwo}%
\providecommand \href [0]{\begingroup \@sanitize@url \@href}%
\providecommand \@href[1]{\@@startlink{#1}\@@href}%
\providecommand \@@href[1]{\endgroup#1\@@endlink}%
\providecommand \@sanitize@url [0]{\catcode `\\12\catcode `\$12\catcode
  `\&12\catcode `\#12\catcode `\^12\catcode `\_12\catcode `\%12\relax}%
\providecommand \@@startlink[1]{}%
\providecommand \@@endlink[0]{}%
\providecommand \url  [0]{\begingroup\@sanitize@url \@url }%
\providecommand \@url [1]{\endgroup\@href {#1}{\urlprefix }}%
\providecommand \urlprefix  [0]{URL }%
\providecommand \Eprint [0]{\href }%
\providecommand \doibase [0]{https://doi.org/}%
\providecommand \selectlanguage [0]{\@gobble}%
\providecommand \bibinfo  [0]{\@secondoftwo}%
\providecommand \bibfield  [0]{\@secondoftwo}%
\providecommand \translation [1]{[#1]}%
\providecommand \BibitemOpen [0]{}%
\providecommand \bibitemStop [0]{}%
\providecommand \bibitemNoStop [0]{.\EOS\space}%
\providecommand \EOS [0]{\spacefactor3000\relax}%
\providecommand \BibitemShut  [1]{\csname bibitem#1\endcsname}%
\let\auto@bib@innerbib\@empty
\bibitem [{\citenamefont {Landau}\ and\ \citenamefont
  {Lifshitz}(1987)}]{Landau-Lifshitz:fluid}%
  \BibitemOpen
  \bibfield  {author} {\bibinfo {author} {\bibfnamefont {L.~D.}\ \bibnamefont
  {Landau}}\ and\ \bibinfo {author} {\bibfnamefont {E.~M.}\ \bibnamefont
  {Lifshitz}},\ }\href@noop {} {\emph {\bibinfo {title} {{Fluid Mechanics}}}},\
  Vol.~\bibinfo {volume} {6}\ (\bibinfo  {publisher} {Butterworth-Heinemann},\
  \bibinfo {address} {Oxford},\ \bibinfo {year} {1987})\BibitemShut {NoStop}%
\bibitem [{\citenamefont {Chaikin}\ and\ \citenamefont
  {Lubensky}(2000)}]{Chaikin:2000}%
  \BibitemOpen
  \bibfield  {author} {\bibinfo {author} {\bibfnamefont {P.~M.}\ \bibnamefont
  {Chaikin}}\ and\ \bibinfo {author} {\bibfnamefont {T.~C.}\ \bibnamefont
  {Lubensky}},\ }\href@noop {} {\emph {\bibinfo {title} {{Principles of
  Condensed Matter Physics}}}}\ (\bibinfo  {publisher} {Cambridge University
  Press, Cambridge, England},\ \bibinfo {year} {2000})\BibitemShut {NoStop}%
\bibitem [{\citenamefont {Rischke}(1996)}]{Rischke:1996}%
  \BibitemOpen
  \bibfield  {author} {\bibinfo {author} {\bibfnamefont {D.~H.}\ \bibnamefont
  {Rischke}},\ }\bibfield  {title} {\bibinfo {title} {Hydrodynamics and
  collective behaviour in relativistic nuclear collisions},\ }\href
  {https://doi.org/https://doi.org/10.1016/S0375-9474(96)00345-4} {\bibfield
  {journal} {\bibinfo  {journal} {Nucle. Phys. A}\ }\textbf {\bibinfo {volume}
  {610}},\ \bibinfo {pages} {88} (\bibinfo {year} {1996})}\BibitemShut
  {NoStop}%
\bibitem [{\citenamefont {Vogelsberger}\ \emph {et~al.}(2014)\citenamefont
  {Vogelsberger}, \citenamefont {Genel}, \citenamefont {Springel},
  \citenamefont {Torrey}, \citenamefont {Sijacki}, \citenamefont {Xu},
  \citenamefont {Snyder}, \citenamefont {Bird}, \citenamefont {Nelson},\ and\
  \citenamefont {Hernquist}}]{Vogelsberger:2014}%
  \BibitemOpen
  \bibfield  {author} {\bibinfo {author} {\bibfnamefont {M.}~\bibnamefont
  {Vogelsberger}}, \bibinfo {author} {\bibfnamefont {S.}~\bibnamefont {Genel}},
  \bibinfo {author} {\bibfnamefont {V.}~\bibnamefont {Springel}}, \bibinfo
  {author} {\bibfnamefont {P.}~\bibnamefont {Torrey}}, \bibinfo {author}
  {\bibfnamefont {D.}~\bibnamefont {Sijacki}}, \bibinfo {author} {\bibfnamefont
  {D.}~\bibnamefont {Xu}}, \bibinfo {author} {\bibfnamefont {G.}~\bibnamefont
  {Snyder}}, \bibinfo {author} {\bibfnamefont {S.}~\bibnamefont {Bird}},
  \bibinfo {author} {\bibfnamefont {D.}~\bibnamefont {Nelson}},\ and\ \bibinfo
  {author} {\bibfnamefont {L.}~\bibnamefont {Hernquist}},\ }\bibfield  {title}
  {\bibinfo {title} {Properties of galaxies reproduced by a hydrodynamic
  simulation},\ }\href {https://doi.org/10.1038/nature13316} {\bibfield
  {journal} {\bibinfo  {journal} {Nature}\ }\textbf {\bibinfo {volume} {509}},\
  \bibinfo {pages} {177} (\bibinfo {year} {2014})}\BibitemShut {NoStop}%
\bibitem [{\citenamefont {Khachatryan}\ \emph {et~al.}()\citenamefont
  {Khachatryan} \emph {et~al.}}]{CMS:2010ifv}%
  \BibitemOpen
  \bibfield  {author} {\bibinfo {author} {\bibfnamefont {V.}~\bibnamefont
  {Khachatryan}} \emph {et~al.} (\bibinfo {collaboration} {CMS
  Collaboration}),\ }\bibfield  {title} {\bibinfo {title} {{Observation of
  long-range near-side angular correlations in proton-proton collisions at the
  LHC}},\ }\href {https://doi.org/10.1007/JHEP09(2010)091} {\bibfield
  {journal} {\bibinfo  {journal} {J. High Energy Phys.}\ }\textbf {\bibinfo
  {volume} {09}},\ \bibinfo {pages} {(2010) 091}}\BibitemShut {NoStop}%
\bibitem [{\citenamefont {Abelev}\ \emph {et~al.}(2013)\citenamefont {Abelev}
  \emph {et~al.}}]{ALICE:2012eyl}%
  \BibitemOpen
  \bibfield  {author} {\bibinfo {author} {\bibfnamefont {B.}~\bibnamefont
  {Abelev}} \emph {et~al.} (\bibinfo {collaboration} {ALICE collaboration}),\
  }\bibfield  {title} {\bibinfo {title} {{Long-range angular correlations on
  the near and away side in $p$-Pb collisions at $\sqrt{s_{NN}}=5.02$ TeV}},\
  }\href {https://doi.org/10.1016/j.physletb.2013.01.012} {\bibfield  {journal}
  {\bibinfo  {journal} {Phys. Lett. B}\ }\textbf {\bibinfo {volume} {719}},\
  \bibinfo {pages} {29} (\bibinfo {year} {2013})}\BibitemShut {NoStop}%
\bibitem [{\citenamefont {Aad}\ \emph {et~al.}(2013)\citenamefont {Aad} \emph
  {et~al.}}]{ATLAS:2012cix}%
  \BibitemOpen
  \bibfield  {author} {\bibinfo {author} {\bibfnamefont {G.}~\bibnamefont
  {Aad}} \emph {et~al.} (\bibinfo {collaboration} {ATLAS collaboration}),\
  }\bibfield  {title} {\bibinfo {title} {{Observation of associated near-side
  and away-side long-range correlations in $\sqrt{s_{NN}}$=5.02 TeV proton-lead
  collisions with the ATLAS detector}},\ }\href
  {https://doi.org/10.1103/PhysRevLett.110.182302} {\bibfield  {journal}
  {\bibinfo  {journal} {Phys. Rev. Lett.}\ }\textbf {\bibinfo {volume} {110}},\
  \bibinfo {pages} {182302} (\bibinfo {year} {2013})}\BibitemShut {NoStop}%
\bibitem [{\citenamefont {Aidala}\ \emph {et~al.}(2019)\citenamefont {Aidala}
  \emph {et~al.}}]{PHENIX:2018lia}%
  \BibitemOpen
  \bibfield  {author} {\bibinfo {author} {\bibfnamefont {C.}~\bibnamefont
  {Aidala}} \emph {et~al.} (\bibinfo {collaboration} {PHENIX collaboration}),\
  }\bibfield  {title} {\bibinfo {title} {{Creation of quark\textendash{}gluon
  plasma droplets with three distinct geometries}},\ }\href
  {https://doi.org/10.1038/s41567-018-0360-0} {\bibfield  {journal} {\bibinfo
  {journal} {Nat. Phys.}\ }\textbf {\bibinfo {volume} {15}},\ \bibinfo {pages}
  {214} (\bibinfo {year} {2019})}\BibitemShut {NoStop}%
\bibitem [{\citenamefont {Abdulhamid}\ \emph {et~al.}(2023)\citenamefont
  {Abdulhamid} \emph {et~al.}}]{STAR:2022pfn}%
  \BibitemOpen
  \bibfield  {author} {\bibinfo {author} {\bibfnamefont {M.~I.}\ \bibnamefont
  {Abdulhamid}} \emph {et~al.} (\bibinfo {collaboration} {STAR
  collaboration}),\ }\bibfield  {title} {\bibinfo {title} {{Measurements of the
  elliptic and triangular azimuthal anisotropies in central $^{3}$He+Au, $d$+Au
  and $p$+Au collisions at $\sqrt{s_{NN}}$=200\,\,GeV}},\ }\href
  {https://doi.org/10.1103/PhysRevLett.130.242301} {\bibfield  {journal}
  {\bibinfo  {journal} {Phys. Rev. Lett.}\ }\textbf {\bibinfo {volume} {130}},\
  \bibinfo {pages} {242301} (\bibinfo {year} {2023})}\BibitemShut {NoStop}%
\bibitem [{\citenamefont {Heller}\ and\ \citenamefont
  {Spali\'{n}ski}(2015)}]{Heller:2015}%
  \BibitemOpen
  \bibfield  {author} {\bibinfo {author} {\bibfnamefont {M.~P.}\ \bibnamefont
  {Heller}}\ and\ \bibinfo {author} {\bibfnamefont {M.}~\bibnamefont
  {Spali\'{n}ski}},\ }\bibfield  {title} {\bibinfo {title} {Hydrodynamics
  beyond the gradient expansion: Resurgence and resummation},\ }\href
  {https://doi.org/10.1103/PhysRevLett.115.072501} {\bibfield  {journal}
  {\bibinfo  {journal} {Phys. Rev. Lett.}\ }\textbf {\bibinfo {volume} {115}},\
  \bibinfo {pages} {072501} (\bibinfo {year} {2015})}\BibitemShut {NoStop}%
\bibitem [{\citenamefont {Kurkela}\ \emph {et~al.}(2019)\citenamefont
  {Kurkela}, \citenamefont {Mazeliauskas}, \citenamefont {Paquet},
  \citenamefont {Schlichting},\ and\ \citenamefont {Teaney}}]{Kurkela:2019}%
  \BibitemOpen
  \bibfield  {author} {\bibinfo {author} {\bibfnamefont {A.}~\bibnamefont
  {Kurkela}}, \bibinfo {author} {\bibfnamefont {A.}~\bibnamefont
  {Mazeliauskas}}, \bibinfo {author} {\bibfnamefont {J.-F.}\ \bibnamefont
  {Paquet}}, \bibinfo {author} {\bibfnamefont {S.}~\bibnamefont
  {Schlichting}},\ and\ \bibinfo {author} {\bibfnamefont {D.}~\bibnamefont
  {Teaney}},\ }\bibfield  {title} {\bibinfo {title} {{Matching the
  nonequilibrium initial stage of heavy ion collisions to hydrodynamics with
  QCD kinetic theory}},\ }\href
  {https://doi.org/10.1103/PhysRevLett.122.122302} {\bibfield  {journal}
  {\bibinfo  {journal} {Phys. Rev. Lett.}\ }\textbf {\bibinfo {volume} {122}},\
  \bibinfo {pages} {122302} (\bibinfo {year} {2019})}\BibitemShut {NoStop}%
\bibitem [{\citenamefont {Romatschke}(2018)}]{Romatschke:2018}%
  \BibitemOpen
  \bibfield  {author} {\bibinfo {author} {\bibfnamefont {P.}~\bibnamefont
  {Romatschke}},\ }\bibfield  {title} {\bibinfo {title} {{Relativistic Fluid
  Dynamics Far From Local Equilibrium}},\ }\href
  {https://doi.org/10.1103/PhysRevLett.120.012301} {\bibfield  {journal}
  {\bibinfo  {journal} {Phys. Rev. Lett.}\ }\textbf {\bibinfo {volume} {120}},\
  \bibinfo {pages} {012301} (\bibinfo {year} {2018})}\BibitemShut {NoStop}%
\bibitem [{\citenamefont {Kurkela}\ \emph {et~al.}(2020)\citenamefont
  {Kurkela}, \citenamefont {van~der Schee}, \citenamefont {Wiedemann},\ and\
  \citenamefont {Wu}}]{kurkela:2020}%
  \BibitemOpen
  \bibfield  {author} {\bibinfo {author} {\bibfnamefont {A.}~\bibnamefont
  {Kurkela}}, \bibinfo {author} {\bibfnamefont {W.}~\bibnamefont {van~der
  Schee}}, \bibinfo {author} {\bibfnamefont {U.~A.}\ \bibnamefont
  {Wiedemann}},\ and\ \bibinfo {author} {\bibfnamefont {B.}~\bibnamefont
  {Wu}},\ }\bibfield  {title} {\bibinfo {title} {Early-and late-time behavior
  of attractors in heavy-ion collisions},\ }\href
  {https://doi.org/10.1103/PhysRevLett.124.102301} {\bibfield  {journal}
  {\bibinfo  {journal} {Phys. Rev. Lett.}\ }\textbf {\bibinfo {volume} {124}},\
  \bibinfo {pages} {102301} (\bibinfo {year} {2020})}\BibitemShut {NoStop}%
\bibitem [{\citenamefont {Romatschke}(2017)}]{Romatschke:2017}%
  \BibitemOpen
  \bibfield  {author} {\bibinfo {author} {\bibfnamefont {P.}~\bibnamefont
  {Romatschke}},\ }\bibfield  {title} {\bibinfo {title} {Relativistic
  hydrodynamic attractors with broken symmetries: non-conformal and
  non-homogeneous},\ }\href {http://dx.doi.org/10.1007/JHEP12(2017)079}
  {\bibfield  {journal} {\bibinfo  {journal} {J. High Energy Phys.}\ }\textbf
  {\bibinfo {volume} {12}},\ \bibinfo {pages} {(2017) 079}}\BibitemShut
  {NoStop}%
\bibitem [{\citenamefont {Florkowski}\ \emph {et~al.}(2018)\citenamefont
  {Florkowski}, \citenamefont {Heller},\ and\ \citenamefont
  {Spali{\'n}ski}}]{Florkowski:2018}%
  \BibitemOpen
  \bibfield  {author} {\bibinfo {author} {\bibfnamefont {W.}~\bibnamefont
  {Florkowski}}, \bibinfo {author} {\bibfnamefont {M.~P.}\ \bibnamefont
  {Heller}},\ and\ \bibinfo {author} {\bibfnamefont {M.}~\bibnamefont
  {Spali{\'n}ski}},\ }\bibfield  {title} {\bibinfo {title} {{New theories of
  relativistic hydrodynamics in the LHC era}},\ }\href
  {https://doi.org/10.1088/1361-6633/aaa091} {\bibfield  {journal} {\bibinfo
  {journal} {Rep. Progr. Phys.}\ }\textbf {\bibinfo {volume} {81}},\ \bibinfo
  {pages} {046001} (\bibinfo {year} {2018})}\BibitemShut {NoStop}%
\bibitem [{\citenamefont {Romatschke}\ and\ \citenamefont
  {Romatschke}(2019)}]{Romatschke:2019}%
  \BibitemOpen
  \bibfield  {author} {\bibinfo {author} {\bibfnamefont {P.}~\bibnamefont
  {Romatschke}}\ and\ \bibinfo {author} {\bibfnamefont {U.}~\bibnamefont
  {Romatschke}},\ }\href@noop {} {\emph {\bibinfo {title} {Relativistic Fluid
  Dynamics In and Out of Equilibrium}}}\ (\bibinfo  {publisher} {Cambridge
  University Press},\ \bibinfo {address} {Cambridge England},\ \bibinfo {year}
  {2019})\BibitemShut {NoStop}%
\bibitem [{\citenamefont {Berges}\ \emph {et~al.}(2021)\citenamefont {Berges},
  \citenamefont {Heller}, \citenamefont {Mazeliauskas},\ and\ \citenamefont
  {Venugopalan}}]{Berges:2021}%
  \BibitemOpen
  \bibfield  {author} {\bibinfo {author} {\bibfnamefont {J.}~\bibnamefont
  {Berges}}, \bibinfo {author} {\bibfnamefont {M.~P.}\ \bibnamefont {Heller}},
  \bibinfo {author} {\bibfnamefont {A.}~\bibnamefont {Mazeliauskas}},\ and\
  \bibinfo {author} {\bibfnamefont {R.}~\bibnamefont {Venugopalan}},\
  }\bibfield  {title} {\bibinfo {title} {{QCD thermalization: Ab initio
  approaches and interdisciplinary connections}},\ }\href
  {https://doi.org/10.1103/RevModPhys.93.035003} {\bibfield  {journal}
  {\bibinfo  {journal} {Rev. Mod. Phys.}\ }\textbf {\bibinfo {volume} {93}},\
  \bibinfo {pages} {035003} (\bibinfo {year} {2021})}\BibitemShut {NoStop}%
\bibitem [{\citenamefont {Soloviev}(2022)}]{Soloviev:2022}%
  \BibitemOpen
  \bibfield  {author} {\bibinfo {author} {\bibfnamefont {A.}~\bibnamefont
  {Soloviev}},\ }\bibfield  {title} {\bibinfo {title} {Hydrodynamic attractors
  in heavy ion collisions: a review},\ }\href
  {http://dx.doi.org/10.1140/epjc/s10052-022-10282-4} {\bibfield  {journal}
  {\bibinfo  {journal} {Eur. Phys. J. C}\ }\textbf {\bibinfo {volume} {82}},\
  \bibinfo {pages} {319} (\bibinfo {year} {2022})}\BibitemShut {NoStop}%
\bibitem [{\citenamefont {Vogt}\ \emph {et~al.}(2012)\citenamefont {Vogt},
  \citenamefont {Feld}, \citenamefont {Fr\"ohlich}, \citenamefont {Pertot},
  \citenamefont {Koschorreck},\ and\ \citenamefont {K\"ohl}}]{Vogt:2012}%
  \BibitemOpen
  \bibfield  {author} {\bibinfo {author} {\bibfnamefont {E.}~\bibnamefont
  {Vogt}}, \bibinfo {author} {\bibfnamefont {M.}~\bibnamefont {Feld}}, \bibinfo
  {author} {\bibfnamefont {B.}~\bibnamefont {Fr\"ohlich}}, \bibinfo {author}
  {\bibfnamefont {D.}~\bibnamefont {Pertot}}, \bibinfo {author} {\bibfnamefont
  {M.}~\bibnamefont {Koschorreck}},\ and\ \bibinfo {author} {\bibfnamefont
  {M.}~\bibnamefont {K\"ohl}},\ }\bibfield  {title} {\bibinfo {title} {{Scale
  invariance and viscosity of a two-dimensional Fermi gas}},\ }\href
  {https://doi.org/10.1103/PhysRevLett.108.070404} {\bibfield  {journal}
  {\bibinfo  {journal} {Phys. Rev. Lett.}\ }\textbf {\bibinfo {volume} {108}},\
  \bibinfo {pages} {070404} (\bibinfo {year} {2012})}\BibitemShut {NoStop}%
\bibitem [{\citenamefont {Brewer}\ and\ \citenamefont
  {Romatschke}(2015)}]{Brewer:2015}%
  \BibitemOpen
  \bibfield  {author} {\bibinfo {author} {\bibfnamefont {J.}~\bibnamefont
  {Brewer}}\ and\ \bibinfo {author} {\bibfnamefont {P.}~\bibnamefont
  {Romatschke}},\ }\bibfield  {title} {\bibinfo {title} {{Nonhydrodynamic
  transport in trapped unitary Fermi gases}},\ }\href
  {https://doi.org/10.1103/PhysRevLett.115.190404} {\bibfield  {journal}
  {\bibinfo  {journal} {Phys. Rev. Lett.}\ }\textbf {\bibinfo {volume} {115}},\
  \bibinfo {pages} {190404} (\bibinfo {year} {2015})}\BibitemShut {NoStop}%
\bibitem [{\citenamefont {Patel}\ \emph {et~al.}(2020)\citenamefont {Patel},
  \citenamefont {Yan}, \citenamefont {Mukherjee}, \citenamefont {Fletcher},
  \citenamefont {Struck},\ and\ \citenamefont {Zwierlein}}]{Patel:2020}%
  \BibitemOpen
  \bibfield  {author} {\bibinfo {author} {\bibfnamefont {P.~B.}\ \bibnamefont
  {Patel}}, \bibinfo {author} {\bibfnamefont {Z.}~\bibnamefont {Yan}}, \bibinfo
  {author} {\bibfnamefont {B.}~\bibnamefont {Mukherjee}}, \bibinfo {author}
  {\bibfnamefont {R.~J.}\ \bibnamefont {Fletcher}}, \bibinfo {author}
  {\bibfnamefont {J.}~\bibnamefont {Struck}},\ and\ \bibinfo {author}
  {\bibfnamefont {M.~W.}\ \bibnamefont {Zwierlein}},\ }\bibfield  {title}
  {\bibinfo {title} {{Universal sound diffusion in a strongly interacting Fermi
  gas}},\ }\href {https://doi.org/10.1126/science.aaz5756} {\bibfield
  {journal} {\bibinfo  {journal} {Science}\ }\textbf {\bibinfo {volume}
  {370}},\ \bibinfo {pages} {1222} (\bibinfo {year} {2020})}\BibitemShut
  {NoStop}%
\bibitem [{\citenamefont {Li}\ \emph {et~al.}(2022)\citenamefont {Li},
  \citenamefont {Luo}, \citenamefont {Wang}, \citenamefont {Xie}, \citenamefont
  {Liu}, \citenamefont {Hu}, \citenamefont {Chen}, \citenamefont {Yao},\ and\
  \citenamefont {Pan}}]{Li:2022}%
  \BibitemOpen
  \bibfield  {author} {\bibinfo {author} {\bibfnamefont {X.}~\bibnamefont
  {Li}}, \bibinfo {author} {\bibfnamefont {X.}~\bibnamefont {Luo}}, \bibinfo
  {author} {\bibfnamefont {S.}~\bibnamefont {Wang}}, \bibinfo {author}
  {\bibfnamefont {K.}~\bibnamefont {Xie}}, \bibinfo {author} {\bibfnamefont
  {X.-P.}\ \bibnamefont {Liu}}, \bibinfo {author} {\bibfnamefont
  {H.}~\bibnamefont {Hu}}, \bibinfo {author} {\bibfnamefont {Y.-A.}\
  \bibnamefont {Chen}}, \bibinfo {author} {\bibfnamefont {X.-C.}\ \bibnamefont
  {Yao}},\ and\ \bibinfo {author} {\bibfnamefont {J.-W.}\ \bibnamefont {Pan}},\
  }\bibfield  {title} {\bibinfo {title} {Second sound attenuation near quantum
  criticality},\ }\href {https://doi.org/10.1126/science.abi4480} {\bibfield
  {journal} {\bibinfo  {journal} {Science}\ }\textbf {\bibinfo {volume}
  {375}},\ \bibinfo {pages} {528} (\bibinfo {year} {2022})}\BibitemShut
  {NoStop}%
\bibitem [{\citenamefont {Li}\ \emph {et~al.}(2024)\citenamefont {Li},
  \citenamefont {Huang},\ and\ \citenamefont {Thomas}}]{Li:2024}%
  \BibitemOpen
  \bibfield  {author} {\bibinfo {author} {\bibfnamefont {X.}~\bibnamefont
  {Li}}, \bibinfo {author} {\bibfnamefont {J.}~\bibnamefont {Huang}},\ and\
  \bibinfo {author} {\bibfnamefont {J.~E.}\ \bibnamefont {Thomas}},\
  }\href@noop {} {\bibinfo {title} {{Universal Hydrodynamic Transport Times in
  the Normal Phase of a Unitary Fermi Gas}}} (\bibinfo {year} {2024}),\ \Eprint
  {https://arxiv.org/abs/2402.14104} {arXiv:2402.14104 [cond-mat.quant-gas]}
  \BibitemShut {NoStop}%
\bibitem [{\citenamefont {Brandstetter}\ \emph {et~al.}(2023)\citenamefont
  {Brandstetter}, \citenamefont {Lunt}, \citenamefont {Heintze}, \citenamefont
  {Giacalone}, \citenamefont {Heyen}, \citenamefont {Ga^^c5^^82ka},
  \citenamefont {Subramanian}, \citenamefont {Holten}, \citenamefont {Preiss},
  \citenamefont {Floerchinger},\ and\ \citenamefont
  {Jochim}}]{Brandstetter:2023}%
  \BibitemOpen
  \bibfield  {author} {\bibinfo {author} {\bibfnamefont {S.}~\bibnamefont
  {Brandstetter}}, \bibinfo {author} {\bibfnamefont {P.}~\bibnamefont {Lunt}},
  \bibinfo {author} {\bibfnamefont {C.}~\bibnamefont {Heintze}}, \bibinfo
  {author} {\bibfnamefont {G.}~\bibnamefont {Giacalone}}, \bibinfo {author}
  {\bibfnamefont {L.~H.}\ \bibnamefont {Heyen}}, \bibinfo {author}
  {\bibfnamefont {M.}~\bibnamefont {Ga^^c5^^82ka}}, \bibinfo {author}
  {\bibfnamefont {K.}~\bibnamefont {Subramanian}}, \bibinfo {author}
  {\bibfnamefont {M.}~\bibnamefont {Holten}}, \bibinfo {author} {\bibfnamefont
  {P.~M.}\ \bibnamefont {Preiss}}, \bibinfo {author} {\bibfnamefont
  {S.}~\bibnamefont {Floerchinger}},\ and\ \bibinfo {author} {\bibfnamefont
  {S.}~\bibnamefont {Jochim}},\ }\href@noop {} {\bibinfo {title} {Emergent
  hydrodynamic behaviour of few strongly interacting fermions}} (\bibinfo
  {year} {2023}),\ \Eprint {https://arxiv.org/abs/2308.09699} {arXiv:2308.09699
  [cond-mat.quant-gas]} \BibitemShut {NoStop}%
\bibitem [{\citenamefont {Fujii}\ and\ \citenamefont
  {Nishida}(2018)}]{Fujii:2018}%
  \BibitemOpen
  \bibfield  {author} {\bibinfo {author} {\bibfnamefont {K.}~\bibnamefont
  {Fujii}}\ and\ \bibinfo {author} {\bibfnamefont {Y.}~\bibnamefont
  {Nishida}},\ }\bibfield  {title} {\bibinfo {title} {Hydrodynamics with
  spacetime-dependent scattering length},\ }\href
  {https://doi.org/10.1103/PhysRevA.98.063634} {\bibfield  {journal} {\bibinfo
  {journal} {Phys. Rev. A}\ }\textbf {\bibinfo {volume} {98}},\ \bibinfo
  {pages} {063634} (\bibinfo {year} {2018})}\BibitemShut {NoStop}%
\bibitem [{\citenamefont {Mori}(1962)}]{Mori:1962}%
  \BibitemOpen
  \bibfield  {author} {\bibinfo {author} {\bibfnamefont {H.}~\bibnamefont
  {Mori}},\ }\bibfield  {title} {\bibinfo {title} {{Collective Motion of
  Particles at Finite Temperatures}},\ }\href
  {https://doi.org/10.1143/PTP.28.763} {\bibfield  {journal} {\bibinfo
  {journal} {Prog. Theor. Exp. Phys.}\ }\textbf {\bibinfo {volume} {28}},\
  \bibinfo {pages} {763} (\bibinfo {year} {1962})}\BibitemShut {NoStop}%
\bibitem [{\citenamefont {Luttinger}(1964)}]{Luttinger:1964}%
  \BibitemOpen
  \bibfield  {author} {\bibinfo {author} {\bibfnamefont {J.~M.}\ \bibnamefont
  {Luttinger}},\ }\bibfield  {title} {\bibinfo {title} {Theory of thermal
  transport coefficients},\ }\href {https://doi.org/10.1103/PhysRev.135.A1505}
  {\bibfield  {journal} {\bibinfo  {journal} {Phys. Rev.}\ }\textbf {\bibinfo
  {volume} {135}},\ \bibinfo {pages} {A1505} (\bibinfo {year}
  {1964})}\BibitemShut {NoStop}%
\bibitem [{\citenamefont {Bradlyn}\ \emph {et~al.}(2012)\citenamefont
  {Bradlyn}, \citenamefont {Goldstein},\ and\ \citenamefont
  {Read}}]{Bradlyn:2012}%
  \BibitemOpen
  \bibfield  {author} {\bibinfo {author} {\bibfnamefont {B.}~\bibnamefont
  {Bradlyn}}, \bibinfo {author} {\bibfnamefont {M.}~\bibnamefont {Goldstein}},\
  and\ \bibinfo {author} {\bibfnamefont {N.}~\bibnamefont {Read}},\ }\bibfield
  {title} {\bibinfo {title} {{Kubo formulas for viscosity: Hall viscosity, Ward
  identities, and the relation with conductivity}},\ }\href
  {https://doi.org/10.1103/PhysRevB.86.245309} {\bibfield  {journal} {\bibinfo
  {journal} {Phys. Rev. B}\ }\textbf {\bibinfo {volume} {86}},\ \bibinfo
  {pages} {245309} (\bibinfo {year} {2012})}\BibitemShut {NoStop}%
\bibitem [{\citenamefont {Fujii}\ and\ \citenamefont
  {Nishida}(2020)}]{Fujii:2020}%
  \BibitemOpen
  \bibfield  {author} {\bibinfo {author} {\bibfnamefont {K.}~\bibnamefont
  {Fujii}}\ and\ \bibinfo {author} {\bibfnamefont {Y.}~\bibnamefont
  {Nishida}},\ }\bibfield  {title} {\bibinfo {title} {{Bulk viscosity of
  resonating fermions revisited: Kubo formula, sum rule, and the dimer and
  high-temperature limits}},\ }\href
  {https://doi.org/10.1103/PhysRevA.102.023310} {\bibfield  {journal} {\bibinfo
   {journal} {Phys. Rev. A}\ }\textbf {\bibinfo {volume} {102}},\ \bibinfo
  {pages} {023310} (\bibinfo {year} {2020})}\BibitemShut {NoStop}%
\bibitem [{\citenamefont {Zwerger}(2012)}]{Zwerger:2012}%
  \BibitemOpen
  \bibinfo {editor} {\bibfnamefont {W.}~\bibnamefont {Zwerger}},\ ed.,\
  \href@noop {} {\emph {\bibinfo {title} {{The BCS--BEC Crossover and the
  Unitary Fermi Gas}}}},\ Lecture Notes in Physics Vol. 836\ (\bibinfo
  {publisher} {Springer},\ \bibinfo {address} {Berlin, Heidelberg},\ \bibinfo
  {year} {2012})\BibitemShut {NoStop}%
\bibitem [{Sup()}]{Suppl}%
  \BibitemOpen
  \href@noop {} {\bibinfo  {journal} {See Supplemental Material, which includes
  the reference~\cite{Aniceto:2019}, for the derivation of the operator version
  of the pressure relation and the derivation of the attractor solution from
  the Borel summation of an expanded solution in the long-time limit}\
  }\BibitemShut {NoStop}%
\bibitem [{\citenamefont {Braaten}\ and\ \citenamefont
  {Platter}(2008)}]{Braaten:2008}%
  \BibitemOpen
\bibfield  {journal} {  }\bibfield  {author} {\bibinfo {author} {\bibfnamefont
  {E.}~\bibnamefont {Braaten}}\ and\ \bibinfo {author} {\bibfnamefont
  {L.}~\bibnamefont {Platter}},\ }\bibfield  {title} {\bibinfo {title} {{Exact
  Relations for a Strongly Interacting Fermi Gas from the Operator Product
  Expansion}},\ }\href {https://doi.org/10.1103/PhysRevLett.100.205301}
  {\bibfield  {journal} {\bibinfo  {journal} {Phys. Rev. Lett.}\ }\textbf
  {\bibinfo {volume} {100}},\ \bibinfo {pages} {205301} (\bibinfo {year}
  {2008})}\BibitemShut {NoStop}%
\bibitem [{\citenamefont {Nishida}\ and\ \citenamefont
  {Son}(2007)}]{Nishida:2007}%
  \BibitemOpen
  \bibfield  {author} {\bibinfo {author} {\bibfnamefont {Y.}~\bibnamefont
  {Nishida}}\ and\ \bibinfo {author} {\bibfnamefont {D.~T.}\ \bibnamefont
  {Son}},\ }\bibfield  {title} {\bibinfo {title} {Nonrelativistic conformal
  field theories},\ }\href {https://doi.org/10.1103/PhysRevD.76.086004}
  {\bibfield  {journal} {\bibinfo  {journal} {Phys. Rev. D}\ }\textbf {\bibinfo
  {volume} {76}},\ \bibinfo {pages} {086004} (\bibinfo {year}
  {2007})}\BibitemShut {NoStop}%
\bibitem [{\citenamefont {Son}(2007)}]{Son:2007}%
  \BibitemOpen
  \bibfield  {author} {\bibinfo {author} {\bibfnamefont {D.~T.}\ \bibnamefont
  {Son}},\ }\bibfield  {title} {\bibinfo {title} {{Vanishing Bulk Viscosities
  and Conformal Invariance of the Unitary Fermi Gas}},\ }\href
  {https://doi.org/10.1103/PhysRevLett.98.020604} {\bibfield  {journal}
  {\bibinfo  {journal} {Phys. Rev. Lett.}\ }\textbf {\bibinfo {volume} {98}},\
  \bibinfo {pages} {020604} (\bibinfo {year} {2007})}\BibitemShut {NoStop}%
\bibitem [{\citenamefont {Enss}(2019)}]{Enss:2019}%
  \BibitemOpen
  \bibfield  {author} {\bibinfo {author} {\bibfnamefont {T.}~\bibnamefont
  {Enss}},\ }\bibfield  {title} {\bibinfo {title} {{Bulk Viscosity and Contact
  Correlations in Attractive Fermi Gases}},\ }\href
  {https://doi.org/10.1103/PhysRevLett.123.205301} {\bibfield  {journal}
  {\bibinfo  {journal} {Phys. Rev. Lett.}\ }\textbf {\bibinfo {volume} {123}},\
  \bibinfo {pages} {205301} (\bibinfo {year} {2019})}\BibitemShut {NoStop}%
\bibitem [{\citenamefont {Dusling}\ and\ \citenamefont
  {Sch\"afer}(2013)}]{Dusling:2013}%
  \BibitemOpen
  \bibfield  {author} {\bibinfo {author} {\bibfnamefont {K.}~\bibnamefont
  {Dusling}}\ and\ \bibinfo {author} {\bibfnamefont {T.}~\bibnamefont
  {Sch\"afer}},\ }\bibfield  {title} {\bibinfo {title} {{Bulk Viscosity and
  Conformal Symmetry Breaking in the Dilute Fermi Gas near Unitarity}},\ }\href
  {https://doi.org/10.1103/PhysRevLett.111.120603} {\bibfield  {journal}
  {\bibinfo  {journal} {Phys. Rev. Lett.}\ }\textbf {\bibinfo {volume} {111}},\
  \bibinfo {pages} {120603} (\bibinfo {year} {2013})}\BibitemShut {NoStop}%
\bibitem [{\citenamefont {Nishida}(2019)}]{Nishida:2019}%
  \BibitemOpen
  \bibfield  {author} {\bibinfo {author} {\bibfnamefont {Y.}~\bibnamefont
  {Nishida}},\ }\bibfield  {title} {\bibinfo {title} {Viscosity spectral
  functions of resonating fermions in the quantum virial expansion},\ }\href
  {https://doi.org/https://doi.org/10.1016/j.aop.2019.167949} {\bibfield
  {journal} {\bibinfo  {journal} {Ann. Phys.}\ }\textbf {\bibinfo {volume}
  {410}},\ \bibinfo {pages} {167949} (\bibinfo {year} {2019})}\BibitemShut
  {NoStop}%
\bibitem [{\citenamefont {Hofmann}(2020)}]{Hofmann:2020}%
  \BibitemOpen
  \bibfield  {author} {\bibinfo {author} {\bibfnamefont {J.}~\bibnamefont
  {Hofmann}},\ }\bibfield  {title} {\bibinfo {title} {High-temperature
  expansion of the viscosity in interacting quantum gases},\ }\href
  {https://doi.org/10.1103/PhysRevA.101.013620} {\bibfield  {journal} {\bibinfo
   {journal} {Phys. Rev. A}\ }\textbf {\bibinfo {volume} {101}},\ \bibinfo
  {pages} {013620} (\bibinfo {year} {2020})}\BibitemShut {NoStop}%
\bibitem [{\citenamefont {Fujii}\ and\ \citenamefont
  {Enss}(2023)}]{Fujii:2023}%
  \BibitemOpen
  \bibfield  {author} {\bibinfo {author} {\bibfnamefont {K.}~\bibnamefont
  {Fujii}}\ and\ \bibinfo {author} {\bibfnamefont {T.}~\bibnamefont {Enss}},\
  }\bibfield  {title} {\bibinfo {title} {Bulk viscosity of resonantly
  interacting fermions in the quantum virial expansion},\ }\href
  {https://doi.org/https://doi.org/10.1016/j.aop.2023.169296} {\bibfield
  {journal} {\bibinfo  {journal} {Ann. Phys.}\ }\textbf {\bibinfo {volume}
  {453}},\ \bibinfo {pages} {169296} (\bibinfo {year} {2023})}\BibitemShut
  {NoStop}%
\bibitem [{\citenamefont {Taylor}\ and\ \citenamefont
  {Randeria}(2012)}]{Taylor:2012}%
  \BibitemOpen
  \bibfield  {author} {\bibinfo {author} {\bibfnamefont {E.}~\bibnamefont
  {Taylor}}\ and\ \bibinfo {author} {\bibfnamefont {M.}~\bibnamefont
  {Randeria}},\ }\bibfield  {title} {\bibinfo {title} {{Apparent low-energy
  scale invariance in two-dimensional Fermi gases}},\ }\href
  {https://doi.org/10.1103/PhysRevLett.109.135301} {\bibfield  {journal}
  {\bibinfo  {journal} {Phys. Rev. Lett.}\ }\textbf {\bibinfo {volume} {109}},\
  \bibinfo {pages} {135301} (\bibinfo {year} {2012})}\BibitemShut {NoStop}%
\bibitem [{\citenamefont {Hofmann}(2011)}]{Hofmann:2011}%
  \BibitemOpen
  \bibfield  {author} {\bibinfo {author} {\bibfnamefont {J.}~\bibnamefont
  {Hofmann}},\ }\bibfield  {title} {\bibinfo {title} {{Current response,
  structure factor and hydrodynamic quantities of a two- and three-dimensional
  Fermi gas from the operator-product expansion}},\ }\href
  {https://doi.org/10.1103/PhysRevA.84.043603} {\bibfield  {journal} {\bibinfo
  {journal} {Phys. Rev. A}\ }\textbf {\bibinfo {volume} {84}},\ \bibinfo
  {pages} {043603} (\bibinfo {year} {2011})}\BibitemShut {NoStop}%
\bibitem [{\citenamefont {Goldberger}\ and\ \citenamefont
  {Khandker}(2012)}]{Goldberger:2012}%
  \BibitemOpen
  \bibfield  {author} {\bibinfo {author} {\bibfnamefont {W.~D.}\ \bibnamefont
  {Goldberger}}\ and\ \bibinfo {author} {\bibfnamefont {Z.~U.}\ \bibnamefont
  {Khandker}},\ }\bibfield  {title} {\bibinfo {title} {Viscosity sum rules at
  large scattering lengths},\ }\href
  {https://doi.org/10.1103/PhysRevA.85.013624} {\bibfield  {journal} {\bibinfo
  {journal} {Phys. Rev. A}\ }\textbf {\bibinfo {volume} {85}},\ \bibinfo
  {pages} {013624} (\bibinfo {year} {2012})}\BibitemShut {NoStop}%
\bibitem [{\citenamefont {G\"otze}\ and\ \citenamefont
  {W\"olfle}(1972)}]{Gotze:1972}%
  \BibitemOpen
  \bibfield  {author} {\bibinfo {author} {\bibfnamefont {W.}~\bibnamefont
  {G\"otze}}\ and\ \bibinfo {author} {\bibfnamefont {P.}~\bibnamefont
  {W\"olfle}},\ }\bibfield  {title} {\bibinfo {title} {Homogeneous dynamical
  conductivity of simple metals},\ }\href
  {https://doi.org/10.1103/PhysRevB.6.1226} {\bibfield  {journal} {\bibinfo
  {journal} {Phys. Rev. B}\ }\textbf {\bibinfo {volume} {6}},\ \bibinfo {pages}
  {1226} (\bibinfo {year} {1972})}\BibitemShut {NoStop}%
\bibitem [{\citenamefont {Hartnoll}\ \emph {et~al.}(2018)\citenamefont
  {Hartnoll}, \citenamefont {Lucas},\ and\ \citenamefont
  {Sachdev}}]{Hartnoll:2018}%
  \BibitemOpen
  \bibfield  {author} {\bibinfo {author} {\bibfnamefont {S.~A.}\ \bibnamefont
  {Hartnoll}}, \bibinfo {author} {\bibfnamefont {A.}~\bibnamefont {Lucas}},\
  and\ \bibinfo {author} {\bibfnamefont {S.}~\bibnamefont {Sachdev}},\
  }\href@noop {} {\emph {\bibinfo {title} {{Holographic Quantum Matter}}}}\
  (\bibinfo  {publisher} {MIT press},\ \bibinfo {address} {Cambridge, MA},\
  \bibinfo {year} {2018})\BibitemShut {NoStop}%
\bibitem [{\citenamefont {Frank}\ \emph {et~al.}(2020)\citenamefont {Frank},
  \citenamefont {Zwerger},\ and\ \citenamefont {Enss}}]{Frank:2020}%
  \BibitemOpen
  \bibfield  {author} {\bibinfo {author} {\bibfnamefont {B.}~\bibnamefont
  {Frank}}, \bibinfo {author} {\bibfnamefont {W.}~\bibnamefont {Zwerger}},\
  and\ \bibinfo {author} {\bibfnamefont {T.}~\bibnamefont {Enss}},\ }\bibfield
  {title} {\bibinfo {title} {{Quantum critical thermal transport in the unitary
  Fermi gas}},\ }\href {https://doi.org/10.1103/PhysRevResearch.2.023301}
  {\bibfield  {journal} {\bibinfo  {journal} {Phys. Rev. Res.}\ }\textbf
  {\bibinfo {volume} {2}},\ \bibinfo {pages} {023301} (\bibinfo {year}
  {2020})}\BibitemShut {NoStop}%
\bibitem [{\citenamefont {Forster}(1975)}]{Forster}%
  \BibitemOpen
  \bibfield  {author} {\bibinfo {author} {\bibfnamefont {D.}~\bibnamefont
  {Forster}},\ }\href@noop {} {\emph {\bibinfo {title} {{Hydrodynamic
  Fluctuations, Broken Symmetry and Correlation Functions}}}}\ (\bibinfo
  {publisher} {WA Benjamin},\ \bibinfo {year} {1975})\BibitemShut {NoStop}%
\bibitem [{\citenamefont {Tan}(2008{\natexlab{a}})}]{Tan:2008a}%
  \BibitemOpen
  \bibfield  {author} {\bibinfo {author} {\bibfnamefont {S.}~\bibnamefont
  {Tan}},\ }\bibfield  {title} {\bibinfo {title} {{Energetics of a strongly
  correlated Fermi gas}},\ }\href
  {https://doi.org/https://doi.org/10.1016/j.aop.2008.03.004} {\bibfield
  {journal} {\bibinfo  {journal} {Ann. Phys. (Amsterdam)}\ }\textbf {\bibinfo
  {volume} {323}},\ \bibinfo {pages} {2952} (\bibinfo {year}
  {2008}{\natexlab{a}})}\BibitemShut {NoStop}%
\bibitem [{\citenamefont {Tan}(2008{\natexlab{b}})}]{Tan:2008b}%
  \BibitemOpen
  \bibfield  {author} {\bibinfo {author} {\bibfnamefont {S.}~\bibnamefont
  {Tan}},\ }\bibfield  {title} {\bibinfo {title} {{Large momentum part of a
  strongly correlated Fermi gas}},\ }\href
  {https://doi.org/https://doi.org/10.1016/j.aop.2008.03.005} {\bibfield
  {journal} {\bibinfo  {journal} {Ann. Phys. (Amsterdam)}\ }\textbf {\bibinfo
  {volume} {323}},\ \bibinfo {pages} {2971} (\bibinfo {year}
  {2008}{\natexlab{b}})}\BibitemShut {NoStop}%
\bibitem [{\citenamefont {Tan}(2008{\natexlab{c}})}]{Tan:2008c}%
  \BibitemOpen
  \bibfield  {author} {\bibinfo {author} {\bibfnamefont {S.}~\bibnamefont
  {Tan}},\ }\bibfield  {title} {\bibinfo {title} {{Generalized virial theorem
  and pressure relation for a strongly correlated Fermi gas}},\ }\href
  {https://doi.org/https://doi.org/10.1016/j.aop.2008.03.003} {\bibfield
  {journal} {\bibinfo  {journal} {Ann. Phys. (Amsterdam)}\ }\textbf {\bibinfo
  {volume} {323}},\ \bibinfo {pages} {2987} (\bibinfo {year}
  {2008}{\natexlab{c}})}\BibitemShut {NoStop}%
\bibitem [{Note1()}]{Note1}%
  \BibitemOpen
  \bibinfo {note} {In Ref.~\cite {Fujii:2018}, the bulk strain rate tensor with
  a space-time dependent scattering length is generally derived as
  $V_{a}(t,\protect \bm {x})=\nabla \cdot \protect \bm {v}(t,\protect \bm
  {x})-3[\partial _t a(t,\protect \bm {x})+\protect \bm {v}(t,\protect \bm
  {x})\cdot \nabla a(t,\protect \bm {x})]/a(t,\protect \bm {x})$, where
  $\protect \bm {v}(t,\protect \bm {x})$ is the fluid velocity and
  $a(t,\protect \bm {x})$ is the space-time dependent scattering length. In
  uniform systems without fluid velocities, the bulk strain rate tensor is
  given only by the time derivative term of the scattering length.}\BibitemShut
  {Stop}%
\bibitem [{\citenamefont {M{\"u}ller}(1967)}]{Muller:1967}%
  \BibitemOpen
  \bibfield  {author} {\bibinfo {author} {\bibfnamefont {I.}~\bibnamefont
  {M{\"u}ller}},\ }\bibfield  {title} {\bibinfo {title} {{Zum Paradoxon der
  W{\"a}rmeleitungstheorie}},\ }\href {https://doi.org/10.1007/BF01326412}
  {\bibfield  {journal} {\bibinfo  {journal} {Z. Phys.}\ }\textbf {\bibinfo
  {volume} {198}},\ \bibinfo {pages} {329} (\bibinfo {year}
  {1967})}\BibitemShut {NoStop}%
\bibitem [{\citenamefont {Israel}(1976)}]{Israel:1976}%
  \BibitemOpen
  \bibfield  {author} {\bibinfo {author} {\bibfnamefont {W.}~\bibnamefont
  {Israel}},\ }\bibfield  {title} {\bibinfo {title} {{Nonstationary
  irreversible thermodynamics: A causal relativistic theory}},\ }\href
  {https://doi.org/https://doi.org/10.1016/0003-4916(76)90064-6} {\bibfield
  {journal} {\bibinfo  {journal} {Ann. Phys. (N.Y.))}\ }\textbf {\bibinfo
  {volume} {100}},\ \bibinfo {pages} {310} (\bibinfo {year}
  {1976})}\BibitemShut {NoStop}%
\bibitem [{\citenamefont {Israel}\ and\ \citenamefont
  {Stewart}(1979)}]{Israel:1979}%
  \BibitemOpen
  \bibfield  {author} {\bibinfo {author} {\bibfnamefont {W.}~\bibnamefont
  {Israel}}\ and\ \bibinfo {author} {\bibfnamefont {J.}~\bibnamefont
  {Stewart}},\ }\bibfield  {title} {\bibinfo {title} {Transient relativistic
  thermodynamics and kinetic theory},\ }\href
  {https://doi.org/https://doi.org/10.1016/0003-4916(79)90130-1} {\bibfield
  {journal} {\bibinfo  {journal} {Ann. Phys. (N.Y.)}\ }\textbf {\bibinfo
  {volume} {118}},\ \bibinfo {pages} {341} (\bibinfo {year}
  {1979})}\BibitemShut {NoStop}%
\bibitem [{Note2()}]{Note2}%
  \BibitemOpen
  \bibinfo {note} {The incomplete Gamma function $\Gamma (-s,-x)$ for $s>0$ and
  $x>0$ appears from the integral $\DOTSI \intop \ilimits@ ^{x}_{x_0}dt\protect
  \tmspace +\thinmuskip {.1667em} e^{t}t^{-s-1}=(-1)^{s+1}[\Gamma
  (-s,-x)-\Gamma (-s,-x_{0})]$ and formally describes the general solution
  independent of the initial conditions for the hydrodynamic equation~\protect
  \textup {\hbox {\mathsurround \z@ \protect \normalfont (\ignorespaces \ref
  {eq:MIS-eq-power-law-drive}\unskip \@@italiccorr )}}.}\BibitemShut {Stop}%
\bibitem [{\citenamefont {Bardon}\ \emph {et~al.}(2014)\citenamefont {Bardon},
  \citenamefont {Beattie}, \citenamefont {Luciuk}, \citenamefont {Cairncross},
  \citenamefont {Fine}, \citenamefont {Cheng}, \citenamefont {Edge},
  \citenamefont {Taylor}, \citenamefont {Zhang}, \citenamefont {Trotzky},\ and\
  \citenamefont {Thywissen}}]{Bardon:2014}%
  \BibitemOpen
  \bibfield  {author} {\bibinfo {author} {\bibfnamefont {A.~B.}\ \bibnamefont
  {Bardon}}, \bibinfo {author} {\bibfnamefont {S.}~\bibnamefont {Beattie}},
  \bibinfo {author} {\bibfnamefont {C.}~\bibnamefont {Luciuk}}, \bibinfo
  {author} {\bibfnamefont {W.}~\bibnamefont {Cairncross}}, \bibinfo {author}
  {\bibfnamefont {D.}~\bibnamefont {Fine}}, \bibinfo {author} {\bibfnamefont
  {N.~S.}\ \bibnamefont {Cheng}}, \bibinfo {author} {\bibfnamefont {G.~J.~A.}\
  \bibnamefont {Edge}}, \bibinfo {author} {\bibfnamefont {E.}~\bibnamefont
  {Taylor}}, \bibinfo {author} {\bibfnamefont {S.}~\bibnamefont {Zhang}},
  \bibinfo {author} {\bibfnamefont {S.}~\bibnamefont {Trotzky}},\ and\ \bibinfo
  {author} {\bibfnamefont {J.~H.}\ \bibnamefont {Thywissen}},\ }\bibfield
  {title} {\bibinfo {title} {{Transverse Demagnetization Dynamics of a Unitary
  Fermi Gas}},\ }\href {https://doi.org/10.1126/science.1247425} {\bibfield
  {journal} {\bibinfo  {journal} {Science}\ }\textbf {\bibinfo {volume}
  {344}},\ \bibinfo {pages} {722} (\bibinfo {year} {2014})}\BibitemShut
  {NoStop}%
\bibitem [{\citenamefont {Luciuk}\ \emph {et~al.}(2017)\citenamefont {Luciuk},
  \citenamefont {Smale}, \citenamefont {B{\"o}ttcher}, \citenamefont {Sharum},
  \citenamefont {Olsen}, \citenamefont {Trotzky}, \citenamefont {Enss},\ and\
  \citenamefont {Thywissen}}]{Luciuk:2017}%
  \BibitemOpen
  \bibfield  {author} {\bibinfo {author} {\bibfnamefont {C.}~\bibnamefont
  {Luciuk}}, \bibinfo {author} {\bibfnamefont {S.}~\bibnamefont {Smale}},
  \bibinfo {author} {\bibfnamefont {F.}~\bibnamefont {B{\"o}ttcher}}, \bibinfo
  {author} {\bibfnamefont {H.}~\bibnamefont {Sharum}}, \bibinfo {author}
  {\bibfnamefont {B.~A.}\ \bibnamefont {Olsen}}, \bibinfo {author}
  {\bibfnamefont {S.}~\bibnamefont {Trotzky}}, \bibinfo {author} {\bibfnamefont
  {T.}~\bibnamefont {Enss}},\ and\ \bibinfo {author} {\bibfnamefont {J.~H.}\
  \bibnamefont {Thywissen}},\ }\bibfield  {title} {\bibinfo {title}
  {{Observation of quantum-limited spin transport in strongly interacting
  two-dimensional Fermi gases}},\ }\href
  {https://doi.org/10.1103/PhysRevLett.118.130405} {\bibfield  {journal}
  {\bibinfo  {journal} {Phys. Rev. Lett.}\ }\textbf {\bibinfo {volume} {118}},\
  \bibinfo {pages} {130405} (\bibinfo {year} {2017})}\BibitemShut {NoStop}%
\bibitem [{\citenamefont {Qi}\ \emph {et~al.}(2021)\citenamefont {Qi},
  \citenamefont {Shi},\ and\ \citenamefont {Zhai}}]{Qi:2021}%
  \BibitemOpen
  \bibfield  {author} {\bibinfo {author} {\bibfnamefont {R.}~\bibnamefont
  {Qi}}, \bibinfo {author} {\bibfnamefont {Z.}~\bibnamefont {Shi}},\ and\
  \bibinfo {author} {\bibfnamefont {H.}~\bibnamefont {Zhai}},\ }\bibfield
  {title} {\bibinfo {title} {Maximum energy growth rate in dilute quantum
  gases},\ }\href {https://doi.org/10.1103/PhysRevLett.126.240401} {\bibfield
  {journal} {\bibinfo  {journal} {Phys. Rev. Lett.}\ }\textbf {\bibinfo
  {volume} {126}},\ \bibinfo {pages} {240401} (\bibinfo {year}
  {2021})}\BibitemShut {NoStop}%
\bibitem [{\citenamefont {Maki}\ \emph {et~al.}(2022)\citenamefont {Maki},
  \citenamefont {Zhang},\ and\ \citenamefont {Zhou}}]{Maki:2022}%
  \BibitemOpen
  \bibfield  {author} {\bibinfo {author} {\bibfnamefont {J.}~\bibnamefont
  {Maki}}, \bibinfo {author} {\bibfnamefont {S.}~\bibnamefont {Zhang}},\ and\
  \bibinfo {author} {\bibfnamefont {F.}~\bibnamefont {Zhou}},\ }\bibfield
  {title} {\bibinfo {title} {{Dynamics of strongly interacting Fermi gases with
  time-dependent interactions: Consequence of conformal symmetry}},\ }\href
  {https://doi.org/10.1103/PhysRevLett.128.040401} {\bibfield  {journal}
  {\bibinfo  {journal} {Phys. Rev. Lett.}\ }\textbf {\bibinfo {volume} {128}},\
  \bibinfo {pages} {040401} (\bibinfo {year} {2022})}\BibitemShut {NoStop}%
\bibitem [{\citenamefont {Du}\ \emph {et~al.}(2021)\citenamefont {Du},
  \citenamefont {Huang},\ and\ \citenamefont {Taya}}]{Du:2021}%
  \BibitemOpen
  \bibfield  {author} {\bibinfo {author} {\bibfnamefont {Z.}~\bibnamefont
  {Du}}, \bibinfo {author} {\bibfnamefont {X.-G.}\ \bibnamefont {Huang}},\ and\
  \bibinfo {author} {\bibfnamefont {H.}~\bibnamefont {Taya}},\ }\bibfield
  {title} {\bibinfo {title} {{Hydrodynamic attractor in a Hubble expansion}},\
  }\href {https://doi.org/10.1103/PhysRevD.104.056022} {\bibfield  {journal}
  {\bibinfo  {journal} {Phys. Rev. D}\ }\textbf {\bibinfo {volume} {104}},\
  \bibinfo {pages} {056022} (\bibinfo {year} {2021})}\BibitemShut {NoStop}%
\bibitem [{Note3()}]{Note3}%
  \BibitemOpen
  \bibinfo {note} {The bulk strain rate tensor coupled to a spatial metric
  $g_{ij}(t,\protect \bm {x})$ is generally given as $V_{a}(t,\protect \bm
  {x})=\DOTSB \sum@ \slimits@ _{k=1}^{3}\nabla _{k}v^{k}(t,\protect \bm
  {x})+\partial _{t}\protect \qopname \relax o{ln}\protect \sqrt {g(t,\protect
  \bm {x})}-3[\partial _{t} a(t,\protect \bm {x})+v^{k}(t,\protect \bm
  {x})\partial _{k} a(t,\protect \bm {x})]/a(t,\protect \bm {x})$, where
  $v^{k}(t,\protect \bm {x})$, $g(t,\protect \bm {x})$, and $a(t,\protect \bm
  {x})$ are fluid velocities, the determinant of the spatial metric, and the
  spacetime-dependent scattering length, respectively~\cite
  {Son:2007,Fujii:2018}. Here, $\nabla _{k}$ is the covariant derivative with
  respect to the spatial metric $g_{ij}(t,\protect \bm {x})$}\BibitemShut
  {NoStop}%
\bibitem [{\citenamefont {Chin}\ \emph {et~al.}(2010)\citenamefont {Chin},
  \citenamefont {Grimm}, \citenamefont {Julienne},\ and\ \citenamefont
  {Tiesinga}}]{Chin:2010}%
  \BibitemOpen
  \bibfield  {author} {\bibinfo {author} {\bibfnamefont {C.}~\bibnamefont
  {Chin}}, \bibinfo {author} {\bibfnamefont {R.}~\bibnamefont {Grimm}},
  \bibinfo {author} {\bibfnamefont {P.}~\bibnamefont {Julienne}},\ and\
  \bibinfo {author} {\bibfnamefont {E.}~\bibnamefont {Tiesinga}},\ }\bibfield
  {title} {\bibinfo {title} {Feshbach resonances in ultracold gases},\ }\href
  {https://doi.org/10.1103/RevModPhys.82.1225} {\bibfield  {journal} {\bibinfo
  {journal} {Rev. Mod. Phys.}\ }\textbf {\bibinfo {volume} {82}},\ \bibinfo
  {pages} {1225} (\bibinfo {year} {2010})}\BibitemShut {NoStop}%
\bibitem [{\citenamefont {Aniceto}\ \emph {et~al.}(2019)\citenamefont
  {Aniceto}, \citenamefont {Ba^^c5^^9far},\ and\ \citenamefont
  {Schiappa}}]{Aniceto:2019}%
  \BibitemOpen
  \bibfield  {author} {\bibinfo {author} {\bibfnamefont {I.}~\bibnamefont
  {Aniceto}}, \bibinfo {author} {\bibfnamefont {G.}~\bibnamefont
  {Ba^^c5^^9far}},\ and\ \bibinfo {author} {\bibfnamefont {R.}~\bibnamefont
  {Schiappa}},\ }\bibfield  {title} {\bibinfo {title} {A primer on resurgent
  transseries and their asymptotics},\ }\href
  {https://doi.org/https://doi.org/10.1016/j.physrep.2019.02.003} {\bibfield
  {journal} {\bibinfo  {journal} {Phys. Rep.}\ }\textbf {\bibinfo {volume}
  {809}},\ \bibinfo {pages} {1} (\bibinfo {year} {2019})}\BibitemShut {NoStop}%
\end{thebibliography}%

\clearpage
\appendix
\pagebreak
\widetext
\begin{center}
\textbf{\large Supplemental Materials:\\
Hydrodynamic Attractor in Ultracold Atoms
}
\end{center}
\setcounter{equation}{0}
\setcounter{figure}{0}
\setcounter{table}{0}
\setcounter{page}{1}
\newcounter{supplementeqcountar}
\newcommand\suppsect[1]{{\it #1.}---}
\renewcommand{\theequation}{S\arabic{equation}}
\renewcommand{\thefigure}{S\arabic{figure}}

\section{Derivation of the complex bulk viscosity as a contact correlation function}
In this section, we derive the expression of the complex bulk viscosity in terms of the contact correlation function.
This section is based on Ref.~\cite{Fujii:2018}.

\subsubsection{The stress tensor operator}
We start with the Hamiltonian density for the two-component fermions with a contact interaction given by
\begin{align}
\label{app-eq:hamiltonian}
\hat{\calH}(t,\bx)
=\sum_{\sigma=\uparrow,\downarrow}
\frac{\bigl(\nabla\hat{\psi}^{\dagger}_{\sigma} (t,\bx)\bigr)\cdot\bigl(\nabla\hat{\psi}_{\sigma} (t,\bx)\bigr)}{2m}
+g_0\hat{\psi}^{\dagger}_{\uparrow}(t,\bx)
\hat{\psi}^{\dagger}_{\downarrow}(t,\bx)
\hat{\psi}_{\downarrow}(t,\bx)
\hat{\psi}_{\uparrow}(t,\bx),
\end{align}
where $\hat{\psi}^{\dagger}_{\sigma}(t,\bx)$ and $\hat{\psi}_{\sigma}(t,\bx)$ are the creation and annihilation operators for a fermion with spin $\sigma$, respectively.
In the dimensional regularization, the coupling constant $g_0$ is renormalized into the $s$-wave scattering length as $mg_0=4\pi a$.
The operators satisfy the equal-time commutation relation,
\begin{align}
[\hat{\psi}_{\sigma}(t,\bx),\,\hat{\psi}_{\tau}^{\dagger}(t,\by)] = \delta_{\sigma\tau}\delta^3(\bx-\by),
\end{align}
and its time evolution is governed by the Heisenberg equation of motion:
\begin{align}
i\pd{}{t}\hat{\psi}_{\sigma}(t,\bx)
& = [\hat{\psi}_{\sigma}(t,\bx),\,\hat{H}(t)] \nonumber \\
&= \biggl[
    -\frac{\nabla^2}{2m}
    +g_0\hat{\psi}^{\dagger}_{\tau}(t,\bx)
    \hat{\psi}_{\tau}(t,\bx)
\biggr]\hat{\psi}_{\sigma}(t,\bx),
\label{app-eq:heisenberg}
\end{align}
with $\hat{H}(t)=\int d^3\bx\,\hat{\calH}(t,\bx)$.

Since the system is invariant under spatial translation, the momentum is conserved.
The corresponding continuity equation follows straightforwardly from the Heisenberg equation of motion as
\begin{align}
 \pd{}{t}\hat{\calP}_i(t,\bx)+\pd{}{x^j}\hat{\Pi}_{ij}(t,\bx)=0,
\end{align}
where the momentum density operator $\hat{\calP}_i(t,\bx)$ and the stress tensor operator $\hat{\Pi}_{ij}(t,\bx)$ are found to be
\begin{align}
\hat{\calP}_{i}(t,\bx)
&\equiv -i\sum_{\sigma=\uparrow,\downarrow}
\hat{\psi}^{\dagger}_{\sigma}(t,\bx)
\tensor{\partial_{i}}\hat{\psi}^{\dagger}_{\sigma}(t,\bx), \\
\hat{\Pi}_{ij}(t,\bx)
&\equiv
\sum_{\sigma=\uparrow,\downarrow}\frac{
    \partial_i\hat{\psi}^{\dagger}_\sigma(t,\bx)
    \partial_j\hat{\psi}_{\sigma}(t,\bx)
    + \partial_j\hat{\psi}^{\dagger}_\sigma(t,\bx)
    \partial_i\hat{\psi}_{\sigma}(t,\bx)
}{2m}
+ \delta_{ij}
    g_0
    \hat{\psi}^{\dagger}_{\uparrow}(t,\bx)
    \hat{\psi}^{\dagger}_{\downarrow}(t,\bx)
    \hat{\psi}_{\downarrow}(t,\bx)
    \hat{\psi}_{\uparrow}(t,\bx)
\end{align}
with $\partial_i\equiv \partial/\partial x^i$ and $\phi\tensor{\partial_i}\varphi=[\phi(\partial_i\varphi)-(\partial_i\phi)\varphi]/2$.
We note that the stress tensor operator is not unique because one can add certain total derivative terms to it while keeping the momentum continuity equation intact.
This ambiguity is irrelevant for our purpose focusing on hydrodynamics~\cite{Fujii:2018}.

By scale variation of the Hamiltonian \eqref{app-eq:hamiltonian} we derive the pressure operator $\hat{p}(t,\bx)\equiv \sum_{i}\hat{\Pi}_{ii}(t,\bx)/3$, which satisfies
\begin{align}
\hat{p}(t,\bx)
=\frac{2}{3}\hat{\calH}(t,\bx)+\frac{\hat{\calC}(t,\bx)}{12\pi m a},
\label{app-eq:operator-pressure-rel}
\end{align}
where $\hat{\calC}(t,\bx)$ is the contact density operator as presented in the main text,
\begin{align}
\hat{\calC}(t,\bx)
\equiv (mg_0)^2 \hat{\psi}^{\dagger}_{\uparrow}(t,\bx)
    \hat{\psi}^{\dagger}_{\downarrow}(t,\bx)
    \hat{\psi}_{\downarrow}(t,\bx)
    \hat{\psi}_{\uparrow}(t,\bx).
\end{align}

\section{Derivation of the attractor solution from the Borel summation}
\subsubsection{Definition of the Borel summation}
We consider a formal power series:
\begin{align}
 A(z)=\sum_{n=0}^{\infty}a_n z^n,
\end{align}
where $a_n$ is a real constant.
Here, $A(z)$ is supposed not necessarily to converge.
For later convenience, we assume that the coefficient $a_{n}$ has a factorial factor proportional to $\Gamma(n+p+1)$ for $p\in \mathbb{R}$, rather than the usual factorial factor $\Gamma(n+1)$.
Let us introduce the Borel transform of $A(z)$ as
\begin{align}
 \calB_{p}[A](s)\equiv\sum_{n=0}^{\infty}\frac{a_n}{\Gamma(n+p+1)}s^{n+p}.
\end{align}
Then, the Borel summation of $A(z)$ is defined as
\begin{align}
 \calS_{p}[A](z)\equiv \textrm{P}\!\!\int^{\infty}_0 ds\,e^{-s}z^{-p}\calB_{p}[A](sz),
\end{align}
where $\textrm{P}$ represents the principal value.
Note that the definitions of the Borel transform and the Borel summation here are slightly extended from the usual definitions with $p=0$ because of the factorial factor $\Gamma(n+p+1)$ of the coefficient $a_{n}$.

The Borel transform $\calB_{p}[A](s)$ of a series $A(z)$ can have singularities at $s>0$.
When a series $A(z)$ is given as a perturbative solution of a certain problem with a small parameter $z$, the singularities of the Borel transform $\calB_{p}[A](s)$ correspond to non-perturbative contributions, which are not captured in the expansion with respect to $z$ such as $e^{-1/z}$~\cite{Aniceto:2019}.
However, such non-perturbative contributions are irrelevant as they correspond to non-hydrodynamic modes for our purposes focusing on hydrodynamic attractors.
Thus, we take the principal value in the definition of the Borel summation $\calS_{p}[A](z)$ as a simple method to avoid the singularities.

\subsubsection{Expanded solution of Eq.~\eqref{eq:MIS-eq-power-law-drive}}
Based on the perspective of gradient expansions in hydrodynamics, we solve Eq.~\eqref{eq:MIS-eq-power-law-drive} by expanding $\pi(t)$ with respect to $\tau_{\zeta}/t$, which is small in the long-time limit.
By factoring out the trivial prefactor $3\zeta[\tilde{a}]\alpha\tau_{\zeta}^{2\alpha}/t^{2\alpha+1}$ corresponding to the right-hand side of Eq.~\eqref{eq:MIS-eq-power-law-drive}, we expand $\pi(t)$ as
\begin{align}
\pi(t)=3\zeta[\tilde{a}]\frac{\alpha\tau_{\zeta}^{2\alpha}}{t^{2\alpha+1}}\tilde{\pi}(\tau_{\zeta}/t),\qquad \tilde{\pi}(z)=\sum_{n=0}^{\infty}\pi_{n}z^{n}.
\label{app-eq:expansion}
\end{align}
Substituting this expansion into Eq.~\eqref{eq:MIS-eq-power-law-drive}, we find
\begin{align}
\pi_{n}=\frac{\Gamma(2\alpha+n+1)}{\Gamma(2\alpha+1)}\qquad \for\quad n=0,1,2,\ldots,
\label{app-eq:coefficient}
\end{align}
where the coefficients are determined independently of the initial condition for $\pi(t)$.
Here, the leading-order solution coincides with the Navier-Stokes hydrodynamic result, and the $n$\,th-order solution gives the $(n+1)$\,th-order hydrodynamic correction.
Importantly, since $\pi_{n}$ is proportional to $\Gamma(2\alpha+n+1)$ and diverges factorially, the expansion~\eqref{app-eq:expansion} does not converge.
Nevertheless, we can obtain a meaningful result for $\pi(t)$ involving information up to infinite order using the aforementioned Borel summation.

\subsubsection{Borel summation of $\pi(t)$}
The Borel transform of $\tilde{\pi}(z)$ is computed as
\begin{align}
\calB_{2\alpha}[\tilde{\pi}](s)
=\sum_{n=0}^{\infty}\frac{\pi_{n}}{\Gamma(n+2\alpha+1)}s^{n+2\alpha}
=\frac{1}{\Gamma(2\alpha+1)}\frac{s^{2\alpha}}{1-s}.
\end{align}
Although the summation in the Borel transform converges only for $|s|<1$, its defined domain can be extended to $s\in\mathbb{C}$ except for $s=1$ by an analytic continuation.
Subsequently, its Borel summation is computed as
\begin{align}
\calS_{2\alpha}[\tilde{\pi}](z)
&= \frac{1}{\Gamma(2\alpha+1)}\textrm{P}\!\!\int^{\infty}_{0}ds\,e^{-s}\frac{s^{2\alpha}}{1-sz}
=e^{-1/z}(-z)^{-2\alpha-1}\Gamma(-2\alpha,-z^{-1}).
\end{align}
Therefore, the corresponding Borel summation of $\pi(t)$ is obtained as
\begin{align}
\pi_{\textrm{Borel}}(t)
=3\zeta[\tilde{a}]\frac{\alpha\tau_{\zeta}^{2\alpha}}{t^{2\alpha+1}}\calS_{2\alpha}[\tilde{\pi}](\tau_{\zeta}/t)
=3\zeta[\tilde{a}]\frac{\alpha}{\tau_{\zeta}}
(-1)^{2\alpha+1}
e^{-t/\tau_{\zeta}}
\Gamma(-2\alpha,-t/\tau_{\zeta}),
\end{align}
which is identical to the attractor solution~\eqref{eq:attractor-sol} in the main text.

\end{document}